\documentclass[a4paper,fleqn,usenatbib]{mnras}

\usepackage[T1]{fontenc}
\usepackage{ae,aecompl}

\usepackage{graphicx}	
\usepackage{amsmath}	
\usepackage{amssymb}	

\usepackage[flushleft]{threeparttable}
\usepackage{afterpage}
\usepackage{multirow}
\usepackage{caption}
\usepackage{lscape}

\title[SCOCENSUS]{DECam Survey for Low-Mass Stars and Substellar Objects in the UCL and LCC Subgroups
of the Sco-Cen OB Association (SCOCENSUS)}

\author[Moolekamp et al.]{
Fred E. Moolekamp$^{1,2}$\thanks{Contact e-mail: \href{mailto:fredem@astro.princeton.edu}{fredem@astro.princeton.edu}}
Eric. E. Mamajek$^{1,3}$
David J. James$^{4,5}$
Kevin L. Luhman$^{6,7}$
\newauthor{
Mark J. Pecaut$^{8}$
Stanmir A. Metchev$^{9,10}$
Cameron P.M. Bell$^{1,11}$
Sara R. Denbo$^{12}$}
\\
$^{1}$Department of Physics and Astronomy, University of Rochester, Rochester, NY 14627\\
$^{2}$Department of Astrophysical Sciences, Princeton University, Princeton, NJ 08544\\
$^{3}$Jet Propulsion Laboratory, California Institute of Technology, 4800 Oak Grove Dr, Pasadena, CA 91109\\
$^{4}$National Optical Astronomy Observatories, Tucson, AZ 85719\\
$^{5}$Department of Astronomy, University of Washington, WA 98195\\
$^{6}$Department of Astronomy and Astrophysics, The Pennsylvania State University, University Park, PA 16802\\
$^{7}$Center for Exoplanets and Habitable Worlds, The Pennsylvania State University, University Park, PA 16802\\
$^{8}$Department of Physics, Rockhurst University, Kansas City, MO 64110\\
$^{9}$Department of Physics \& Astronomy and Centre for Planetary Science and Exploration,\\
The University of Western Ontario, London, Ontario, Canada N6A 3K7\\
$^{10}$Department of Physics \& Astronomy, Stony Brook University, Stony Brook, NY 11794-3800\\
$^{11}$Leibniz Institute for Astrophysics, Potsdam (AIP), An der Sternwarte 16, 14482 Potsdam, Germany\\
$^{12}$Department of Physics and Astronomy, Michigan State University, East Lansing, MI 48824}

\date{Accepted XXX. Received YYY; in original form ZZZ}

\pubyear{2017}

\begin{document}
\label{firstpage}
\pagerange{\pageref{firstpage}--\pageref{lastpage}}
\maketitle

\begin{abstract}
Using images taken with the Dark Energy Camera (DECam), the
first extensive survey of low mass and substellar objects is made in
the 15-20 Myr Upper Centaurus Lupus (UCL) and Lower Centaurus Crux (LCC) subgroups of the Scorpius Centaurus OB Association (Sco-Cen).
Due to the size of our dataset (>2Tb) we developed an extensive open source set of python
libraries to reduce our images, including astrometry, coaddition, and
PSF photometry. Our survey consists of 29$\times$3 deg$^2$ fields in
the UCL and LCC subgroups of Sco-Cen and the creation of a catalog
with over 11 million point sources. We create a prioritized list of
candidate for members in UCL and LCC, with 118 \emph{best} and
another 348 \emph{good} candidates. We show that the luminosity and
mass functions of our low mass and substellar candidates are
consistent with measurements for the younger Upper Scorpius subgroup
and estimates of a universal IMF, with spectral types ranging from
M1 down to L1.
\end{abstract}




\section{Introduction} \label{scoii-intro}

Obtaining well-characterized samples of young low-mass objects
(YLMOs), which includes low-mass stars and brown dwarfs, is needed for modeling the early evolution
of low mass stars and substellar objects.
This includes understanding how their circumstellar disks evolve,
observationally constraining the initial mass function, and
determining whether or not a separate formation mechanism is required to
explain the formation of the lowest-mass objects.
While spectral indicators of low surface gravity have been used to identify M and L
dwarfs with large radii, likely due to youth \citep{Kirkpatrick2008,
Cruz2009, Faherty2013}, a method for determining accurate ages of
field YLMOs has remained elusive. The most successful method to age
date YLMOs is to establish their membership in a young cluster or
association, where the age of the stellar members can be used to
estimate the age of the low mass stellar and substellar population
\citep{Luhman2000, Slesnick2004, Bihain2010, Sung2010}.
While many young clusters have been surveyed down to the hydrogen burning limit,
some even down to the deuterium burning limit,
the number of substellar objects with ages between 10-20 Myr remains very limited,
leaving a vital gap in our knowledge of star and planet formation.

One of the most important indirect observables of star formation is
the initial mass function (IMF). The power-law that describes stellar
objects with $m\geq 1 M_\odot$ has not significantly changed since
\citet{Salpeter1955}, which described the IMF as
$\Phi\left(\log(m)\right)=Am^{-\Gamma}$, where
$\Gamma\sim1.35$ \citep{Chabrier2003, Lada2003, Bastian2010, Krumholz2014, Dib2014}.
However, the characteristic mass that defines the
turnover between 0.1-0.3 $M_\odot$ and the shape of the low mass
stellar and substellar IMF is still under investigation
\citep{Bastian2010}. Distinguishing between the most popular fits of
the IMF: the power-law of \citet{Kroupa2002}, the log-normal form of
\citet{Chabrier2003}, and the smoother tapered power-law of
\citet{deMarchi2005}; requires observations of substellar objects
below 40-50 $M_\textrm{Jup}$.
Part of the difficulty in measuring the IMF is that even when
the mass of a population of stars and substellar objects is well constrained,
what is measured is the present day mass function (PDMF),
the current distribution of masses in an observed region.
Often what is desired is to measure the \emph{creation function},
the number of stars per unit volume that form in an infinitesimally small
range or an infinitely small time interval \citep{Miller1979}.
In order to minimize the number of stars that have been ejected or kinematically evolved
from a star-forming region,
and the number of higher mass stars that have evolved off the
main sequence and into end states of stellar evolution that are more difficult to observe,
it is useful to observe young clusters and associations whose members can be
identified by similar positions and kinematics \citep{Chabrier2003, Lada2003, Reid2005, Bastian2010, Krumholz2014}.

Sco-Cen, the nearest OB association to the solar system, is comprised
of the Upper Scorpius (Upper Sco), Upper Centaurus Lupus (UCL), and
Lower Centaurus Crux (LCC) subgroups, with mean ages of
$\sim$10, $\sim$16, and $\sim$16 Myr \citep{Preibisch2008, Pecaut2012,
Pecaut2016}.  Studies of the stellar and substellar populations of
Sco-Cen have played an important role in our understanding of the
evolution of star-forming regions, where research over the past decade
has brought our picture of star formation in the region into focus
\citep{Hoogerwerf2001, Chatterjee2004, Preibisch2008, Feiden2016,
Pecaut2016}.  Covering nearly 2000 deg${}^2$, much of the stellar
population of Sco-Cen was detected by \citet{deZeeuw1999} and earlier
surveys of OB stars.  Efforts to find low mass and substellar members
have been focused on the younger and more dense Upper Sco subgroup
\citep{Ardila2000}, with very few M dwarfs found in the slightly older
UCL and LCC subgroups.  Given the typical ratio of $\sim$6 stars for
every brown dwarf discovered in nearby star-forming clouds
\citep{Luhman2012}, each square degree of UCL and LCC are likely to have
only one or two substellar members and fewer than a dozen low mass
stars, requiring a large survey to detect a statistically
significant sample.  For this reason we utilized the Dark Energy
Camera (DECam, \citealt{DePoy2008}), an optical camera mounted on
the prime focus of the Blanco 4-m telescope at Cerro Tololo with a 3
deg${^2}$ field of view, capable of detecting $\sim$3 substellar
objects per field. By combining DECam $izY$ photometry and astrometry
with all-sky catalogs we were able to detect hundreds of new Sco-Cen
candidates, including a few dozen objects likely to be substellar (see
Section \ref{scoii-imf} for more on the complications of estimating mass for our
candidates).

Until recently, stellar evolution models have tended to
under-predict both the ages of lower mass stars and their
luminosities \citep{Bell2012, Pecaut2012, Kraus2015, David2016,
Pecaut2016, Rizzuto2016}.  There were many theories regarding the
cause of this discrepancy, including an improved understanding of
opacities in low mass stellar atmospheres due to the formation of
molecules \citep{Hillenbrand2008}, but the recent results of
\citet{Feiden2016} suggest that magnetic inhibition of convection
causes the largest deviation between observations of K and M stars and
the predictions of earlier models like \citet{Baraffe2015}.  While
this has improved our understanding of pre-main sequence stellar
evolution, the same cannot be said for their lower mass counterparts.
\citet{Feiden2016} has proven quite successful in predicting the
properties of known members of Upper Scorpius, however the models are
limited to stellar sources above 85 M${}_\textrm{Jup}$, leaving young
low mass stars near the hydrogen burning limit and substellar objects without
evolutionary models that match observations.

Observations of young clusters and associations are also useful in estimating the
timescale in which gaseous disks dissipate, which is mass dependent and presently under debate.
For example, \citet{Carpenter2006} and \citet{Luhman2012} both investigated $\sim$10 Myr Upper Sco,
with \citet{Carpenter2006} not finding any evidence for accretion disks in their sample of F and G stars
and \citet{Luhman2012} estimating that $<$10\% of their B-G stars showed evidence of accretion.
However, both surveys found an increased fraction of gaseous disks in K \citep{Carpenter2006} and
M \citep{Carpenter2006, Luhman2012} stars in the same subgroup.
A similar survey of F stars in the $\sim$5 Myr $\lambda$ Ori cluster by \citet{Hernandez2009}
also showed that accretion disks had all but vanished,
while a number of F stars still displayed ``moderate'' 24$\mu$m excess,
indicating the presence of debris disks.
But mounting evidence is suggesting that lower mass stars and brown dwarfs can have
accretion disks with much longer lifetimes.
\citet{Barrado2007} performed a survey of K- and M-type stars and found that
accretion rates increased with later type stars, which could be indicative of gaseous disks.
\citet{Reiners2009} discovered a brown dwarf kinematically consistent with the $\sim$45
Myr-old Tuc-Hor association \citep[2MASS J004135.39-562112.77; age
from ][]{Bell2015}. They estimate the mass of the brown dwarf to be
under 60 M${}_{Jup}$ and as low as 35 M${}_{Jup}$, at an age of
$\gtrsim$15 Myr, indicating that accretion disks around substellar
objects can last significantly longer that they do around higher mass stars.
More recently, the discovery of an accretion disk around a 0.1 M$_\odot$ star in the 45 Myr
Carina association by \citet{Murphy2017} indicates that even low mass stars might have
longer-lived accretion disks than previously thought.

Once the gas has dissipated, the lifetime of the remaining debris disk is also not well understood.
\citet{Hernandez2009} also investigated the fraction of stars with disks
(disk fraction) in $\lambda$ Ori and several other groups and determined that
while lower mass stars did indeed appear to keep their primordial accretion disks longer,
intermediate mass stars (B0-F0) had a much higher debris disk fraction.
They also noticed that the disk fraction for intermediate mass stars appeared to \emph{increase} with age, from $\sim$20\% at $\sim$3 Myr, to $\sim$40\% at $\sim$5 Myr, up to $\sim$50\% at $\sim$10 Myr.
They hypothesize that the explanation for the contrast in the debris disk fraction
between low and intermediate mass stars, as well as the increased intermediate mass disk fraction with age,
can be explained by a second generation of dust making a substantial contribution to the
thickness of the disk.
Understanding the fraction of YLMOs with debris disks for a wider range of masses in Sco-Cen
will provide valuable insight into the later stages of disk evolution.

This paper presents the preliminary results of SCOCENSUS, a survey of
UCL and LCC for low mass and substellar objects using DECam.  This
work includes a catalog of over 11 million point sources and a
smaller catalog of photometrically and kinematically selected
candidate members of the two subgroups.  Future follow-up spectroscopy
will be used to confirm membership and explore the properties of our
candidates. Section \ref{scoii-obs} describes our observations,
Section \ref{scoii-candidates} discusses our procedure for selecting
candidate members, Section \ref{scoii-lfimf} shows our calculation of
the luminosity function of UCL and LCC and gives a preliminary estimation
of the initial mass function, Section \ref{scoii-disks} investigates
the presence of disks in our candidates, and Section \ref{scoii-etacha}
discusses our null result in detecting substellar members in the
$\eta$-Cha complex.

\section{Observations}\label{scoii-obs}

\begin{figure}
    \includegraphics[width=.48\textwidth]{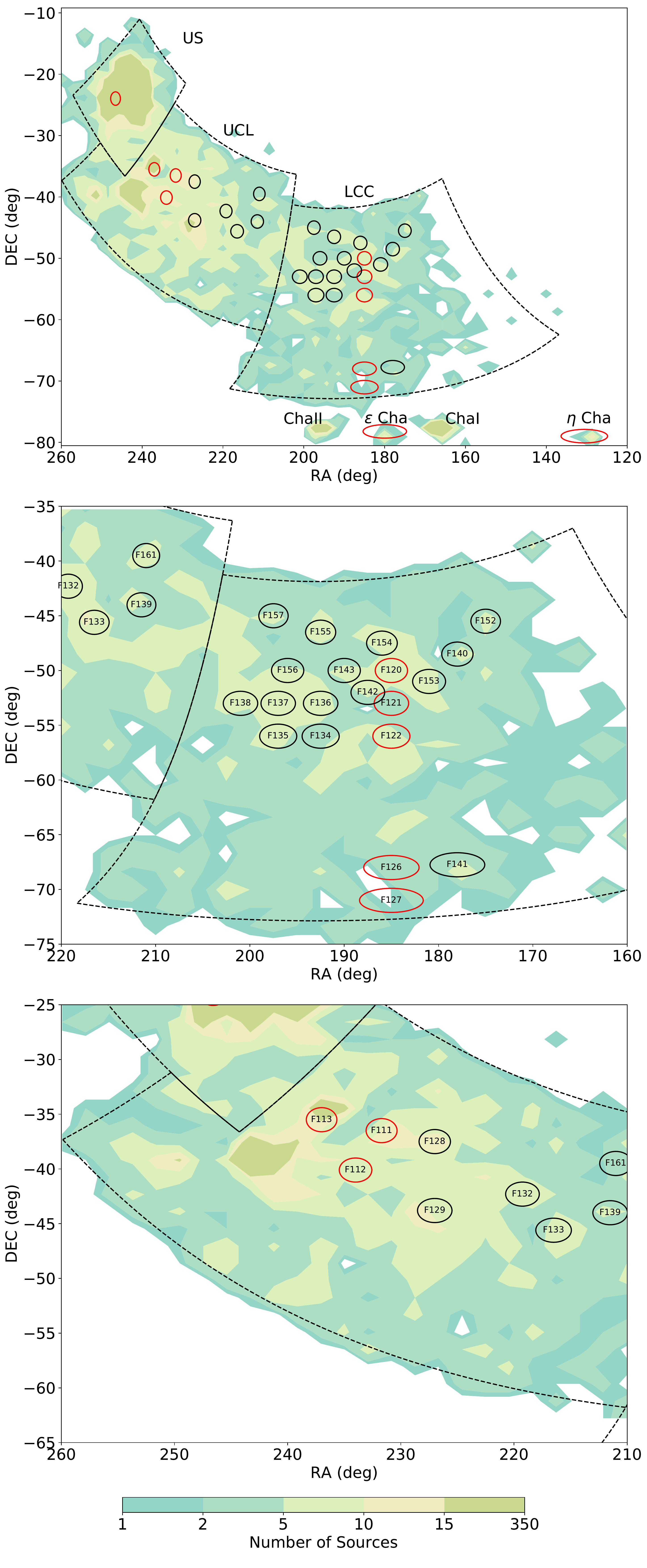}
    \caption{Density map of the known (stellar) Sco-Cen members in DECam fields
      observed in May 2015. Red circles mark fields observed in both
      2013 and 2015 while black circles were only observed in 2015. The
      top plot shows all of Sco-Cen and the fields we observed while
      the middle and lower panels show UCL and LCC respectively, along with the names
      of the fields used in our survey (field names populate the ``OBJECT'' key in the
      DECam FITS headers and are listed for each source in our catalog).}\label{fig:all_fields}
\end{figure}

\afterpage{
    \clearpage
    \begin{landscape}
    \scriptsize
    \begin{threeparttable}
    \caption{DECam observations of low mass stellar and substellar candidates in UCL and LCC}\label{tab:candidates}

    \begin{tabular}{ccccccccccccc}
    \hline \hline
    id & ra & dec & i & z & Y & $\mu_{\alpha_D}$ & $\mu_{\delta_D}$ & $\mu_{\alpha_s}$ & $\mu_{\delta_s}$ & subgroup & confidence & SCOCENSUS \\
     &  &  & (mag) & (mag) & (mag) & (mas yr${}^{-1}$) & (mas yr${}^{-1}$) & (mas yr${}^{-1}$) & (mas yr${}^{-1}$) &  &  &  \\
    \hline
    Moolekamp 1 & 12:17:47.649 & -49:15:49.279 & $14.80\pm 0.01$ & $13.93\pm 0.01$ & $12.81\pm 0.01$ & $-36.7\pm 4.91$ & $-8.41\pm 2.31$ & $-29.8\pm 2.55$ & $-14.6\pm 2.72$ & LCC & best & 120-6004 \\
    Moolekamp 2 & 12:15:51.238 & -49:37:34.889 & $17.35\pm 0.01$ & $15.97\pm 0.01$ & $14.68\pm 0.02$ & $-27.5\pm 4.30$ & $-15.4\pm 4.43$ & $-36.0\pm 3.21$ & $-10.9\pm 1.68$ & LCC & best & 120-20608 \\
    Moolekamp 3 & 12:24:23.400 & -49:38:44.179 & $14.74\pm 0.01$ & $13.77\pm 0.02$ & $12.65\pm 0.01$ & $-30.1\pm 5.87$ & $-14.8\pm 5.77$ & $-31.2\pm 3.49$ & $-20.6\pm 3.46$ & LCC & best & 120-27849 \\
    Moolekamp 4 & 12:20:50.185 & -49:56:58.484 & $16.29\pm 0.01$ & $15.25\pm 0.01$ & $14.20\pm 0.01$ & $-26.4\pm 5.30$ & $-4.69\pm 7.11$ & $-25.2\pm 3.48$ & $-15.3\pm 3.54$ & LCC & best & 120-45885 \\
    Moolekamp 5 & 12:16:42.185 & -50:00:37.962 & $16.50\pm 0.01$ & $15.34\pm 0.01$ & $14.17\pm 0.01$ & $-31.2\pm 5.12$ & $-7.85\pm 6.83$ & $-40.9\pm 3.46$ & $-31.5\pm 3.56$ & LCC & good & 120-52544 \\
    \hline
    \end{tabular}
    \footnotesize
    \begin{tablenotes}
    \item All coordinates are given from the May 2015 observations. $\mu_{\alpha_D}$ and $\mu_{\delta_D}$ denote the \emph{DECam} proper motions while $\mu_{\alpha_s}$ and $\mu_{\delta_s}$ are the \emph{Sky} proper motions (see Section \ref{scoii-obs}). SCOCENSUS is the survey id used in the Vizier catalog (SCOCENSUS NNN-NNNNNN) for each source while id is a special identifier for all of the candidates described in this paper. Only the first five rows are shown; this table is available in its entirety in the electronic version of the journal\end{tablenotes}
    \end{threeparttable}
    \end{landscape}
    \clearpage
}

Our survey uses observations taken with DECam, mounted at the prime focus of
the Blanco 4-m telescope at Cerro Tololo Inter-American Observatory
(CTIO). DECam has 60 or 61 2048$\times$4096 functioning CCDs
(depending on the observation date) and a field of view (FOV) of
$\sim$3 deg${}^2$ (2.2 deg wide) with a 0.263 arcsec pixel${}^{-1}$
resolution. Gaps between the rows and columns of CCDs are 201 pix
($\sim$53 arcsec) and 153 pix ($\sim$40.3 arcsec) respectively.
The images were taken in the $izY$ filters (775, 925, 1000 nm central wavelengths
respectively), where the $iz$ filters were designed to closely match
SDSS $iz$ (762 nm, 913 nm respectively).
Our images were taken during two separate runs: the first run
on 2013 May 29-30, which was only photometric for 5 hours and a second
run on 2015 May 26-27. We observed a total of 6 photometric fields in
2013 and 29 in 2015 (see Figure \ref{fig:all_fields}). Each science
image was observed in each band for 7s, then 3 exposures using the
same pointing in each filter as follows: 200s in $i$ and 30s in
$zY$. For photometric calibration we also observed three SDSS fields centered at:
(14:42:00,-00:04:50), (12:27:00,-00:04:50), (10:48:00,00:00:10). The
calibration fields were observed once every 1-2 hours for 30s in $i$
and 15s in $zY$. The mean completeness limits for each filter are
$i=21.8$, $z=20.4$, $Y=18.6$ mag with saturation at $i=12.6$,
$z=12.4$, and $Y=11.6$ mag.

The DECam images were reduced using the steps described in more detail in Appendix
\ref{scocensus}. Briefly, using the InstCal images produced by the
DECam community Pipeline \citep{Valdes2014}, we stack the longer
exposures in $izY$ individually for each CCD and combine the catalogs
generated from the stacks with the 7s exposure catalogs in all three
filters to create a master catalog of point sources. PSF photometry is
performed on the individual CCDs and calibrated to the SDSS AB system
(for $iz$) and UKIDSS Vega system (for $Y$) using the standard
fields. An astrometric solution is derived for each CCD in 2013 and
2015 and the 6 fields observed in both epochs are compared to generate
relative proper motions that are calibrated to \emph{Gaia} observations
(referred to as \emph{DECam} proper motions for the remainder of this
paper). All of the fields also have proper motions calculated by
matching our catalog to the 2MASS \citep{Skrutskie2006}, DENIS
\citep{Epchtein1997}, AllWISE \citep{Wright2010}, UCAC4
\citep{Zacharias2013}, USNOB 1.0 \citep{Monet2003}, and
GAIA \citep{GAIA2016} catalogs
(referred to as \emph{Sky} proper motions for the remainder of this
paper). Fields observed in both epochs have both \emph{DECam} and
\emph{Sky} proper motions recorded in the final catalog.

\section{SCOCENSUS Catalogs}\label{scoii-candidates}

The total number of point sources detected in the 2013 and 2015 DECam
imagery was over 11 million, with $\sim$5 million sources detected in
both $i$ and $z$ and $\sim$3 million detected in both $z$ and $Y$. For
each source the full catalog contains the epoch, position, DECam $izY$
photometry, 2MASS $JHK_s$ photometry (when available), AllWISE
photometry (when available), DENIS $IJK$ photometry (when available),
\emph{Sky} proper motions (see previous section), \emph{DECam} proper
motions (when available, see previous section), the name of the
observed field, Sco-Cen subgroup of the field (Upper Sco, UCL, LCC),
and various flags created in the execution of the pipeline. The full
catalog is available on Vizier.

Our fields were taken at a variety of Galactic latitudes, causing the
color-magnitude diagrams (CMDs) to suffer from increased reddeding
at lower Galactic latitudes. In addition to the spatial separation between UCL
and LCC, the clustering of ages illustrated in Figure 9 from
\citet{Pecaut2016} results in potentially different distances and ages
for all of our DECam fields. As a result we perform our candidate
selection on each field individually to select candidates and
calculate their estimated properties.

\subsection{Candidates}

Candidate members in UCL and LCC are selected using photometric and kinematic cuts
that appear to have a false positive rate of $\sim5-6\%$.
We first use our DECam photometry to select objects near isochrones created using the
\citet{Baraffe2015} stellar evolution models calibrated with \citet{Allard2011} atmospheric models,
eliminating all but $4.7\times10^{-2}$\% of the sources in our catalog.
Next we select objects with proper motions consistent with \citet{Chen2011} estimates of
UCL and LCC group velocities, dividing our candidates into ``best'', ``good'', and ``no pm''
categories based on their observed proper motions.
Finally we use 2MASS photometry to remove M giants with similar $izY$ colors that occupy
a different region in ($H-K_s$,$J-H$) color-color space than YLMOs \citep{Cruz2009}.
The definition of our proper motion categories, a more thorough explanation of 
how we built our models and the procedure used to select candidates is described in
Appendix \ref{scoii-select-members}.

Once all of our cuts have been made, we are left with 118 \emph{best} candidates in the fields observed in
2013 and 2015 and 348 more \emph{good} candidates as defined in Appendix \ref{scoii-select-members}.
While our full catalog of candidates contains
many more fields, we present a few tables with the most important
fields for each candidate. Table \ref{tab:candidates} lists the observed properties
for all of the candidates.

\begin{figure*}\centering
    \includegraphics[width=1.0\textwidth]{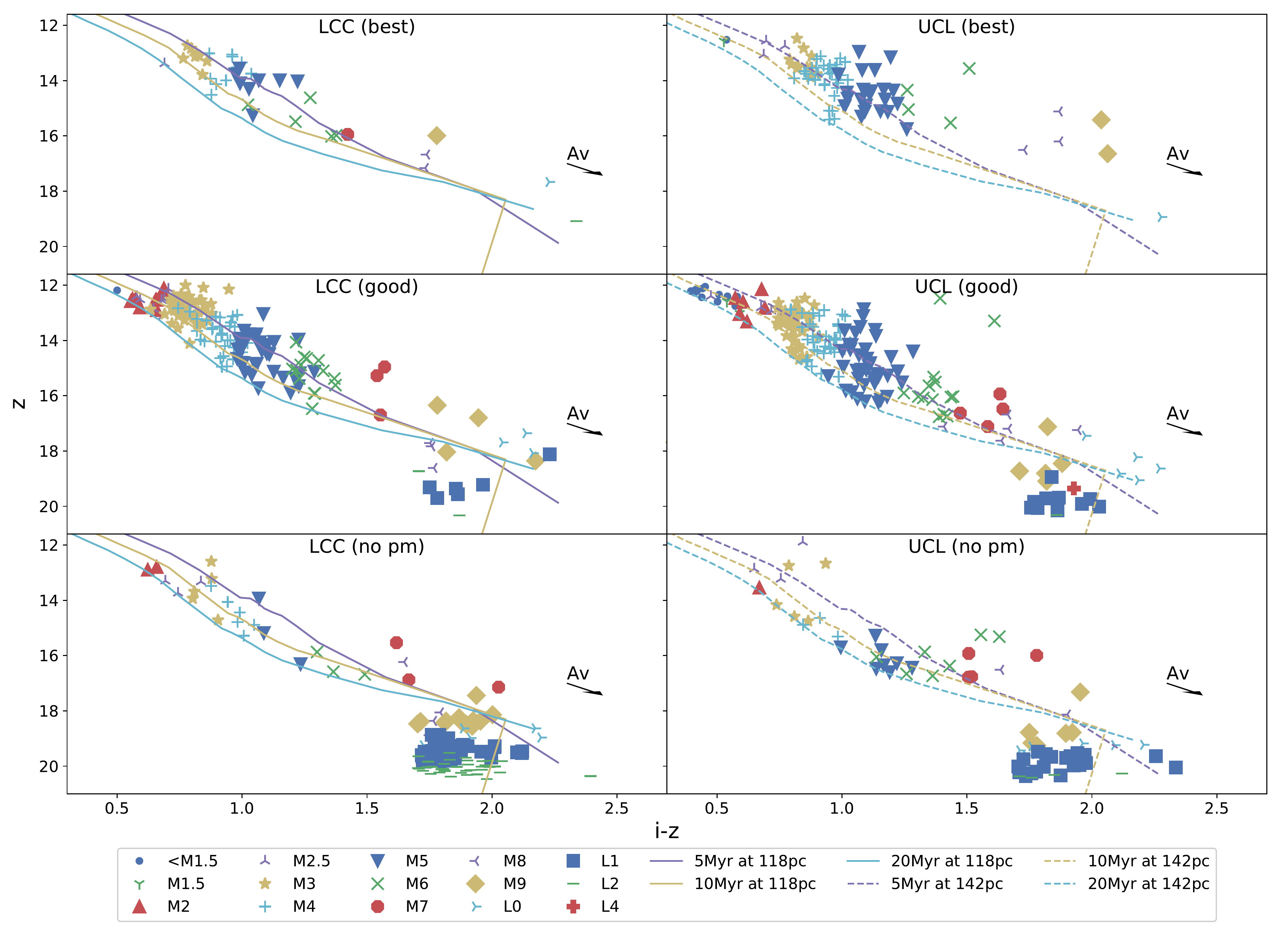}
    \caption{YLMO candidates in LCC (left) and UCL (right).
      Error bars have been omitted for clarity, as the errors for most
      sources are smaller than the size of the markers.
      The top row shows the sources observed in two epochs with
      proper motions and photometry consistent with membership in
      Sco-Cen. The middle row shows sources observed in only 2015
      with proper motions and photometry consistent with Sco-Cen
      membership, and the bottom row shows objects observed in 2015
      that are photometrically consistent with Sco-Cen membership but
      are not found in any existing all-sky catalog, and thus have no
      proper motion estimates. Due to a lack of proper motions and
      limited photometric bands (missing $JHK_s$ photometry makes it more
      difficult to rule out giants), most of
      the objects in the bottom row are expected to be non-members.
      See Appendix \ref{scoii-select-members} for further information.}
    \label{fig:lcc_ucl_candidates}
\end{figure*}

As described in Appendix \ref{scoii-select-members}, we also derived
estimates for the effective temperature, extinction, spectral type,
and mass for all of the candidates. Table \ref{tab:derived} shows the derived
properties for all of the candidates in UCL and LCC.
We use lower case letters and the heading ``Spectral
Template'' to make it clear that our spectral types are photometric
estimates and not based on observed spectra (however in Section
\ref{scoii-etacha} we show that our estimates appear to be within a
subtype for all candidates that have observed spectra).

\begin{table}
\caption{Derived properties of low mass stellar and substellar candidates in UCL and LCC}\label{tab:derived}
\begin{threeparttable}
\begin{tabular}{ccccc}
\hline \hline
id & $T_{eff}$ & Av & Spectral & Mass \\
 & (K) &  & Template & ($M_{\odot}$) \\
\hline
Moolekamp 1 & $3163^{+103}_{-103}$ & $0.16^{+0.19}_{-0.11}$ & m4 & $0.16^{+0.04}_{-0.05}$ \\
Moolekamp 2 & $2763^{+73}_{-73}$ & $0.15^{+0.19}_{-0.10}$ & m6 & $0.05^{+0.01}_{-0.01}$ \\
Moolekamp 3 & $3106^{+92}_{-92}$ & $0.15^{+0.20}_{-0.09}$ & m5 & $0.13^{+0.06}_{-0.03}$ \\
Moolekamp 4 & $3081^{+71}_{-71}$ & $0.19^{+0.19}_{-0.12}$ & m5 & $0.11^{+0.04}_{-0.01}$ \\
Moolekamp 5 & $2959^{+72}_{-72}$ & $0.17^{+0.18}_{-0.11}$ & m5 & $0.07^{+0.02}_{-0.01}$ \\
\hline
\end{tabular}
\begin{tablenotes}[flushleft]
\item Only the first five rows are shown; this table is available in its entirety in the electronic version of the journal\end{tablenotes}
\end{threeparttable}
\end{table}

\section{Luminosity and Mass Functions} \label{scoii-lfimf}

\subsection{Luminosity Functions in UCL and LCC}\label{scoii-lf}

Before calculating the IMF it can be useful to first estimate the
luminosity function (LF), which describes the distribution of absolute
magnitudes of the observed cluster or association (the term
``luminosity function'' is a bit misleading as historically it has
been a distribution and not a function that is calculated). For much
of the 20th century, the calculation of absolute magnitudes for low
mass stars in the field proved to be quite challenging (distance
estimates based on parallax were accurate to $\lesssim$10 pc), leading
to much disagreement and controversy regarding the abundance of M
dwarfs and the shape of the LF \citep{Reid2005}. For young clusters
and associations, both the distance to the groups and approximate
reddening are known from astrometric and spectral observations of
their stellar populations and make the conversion of apparent to
absolute magnitudes somewhat simpler.

\begin{figure}\centering
    \includegraphics[width=.3\textwidth]{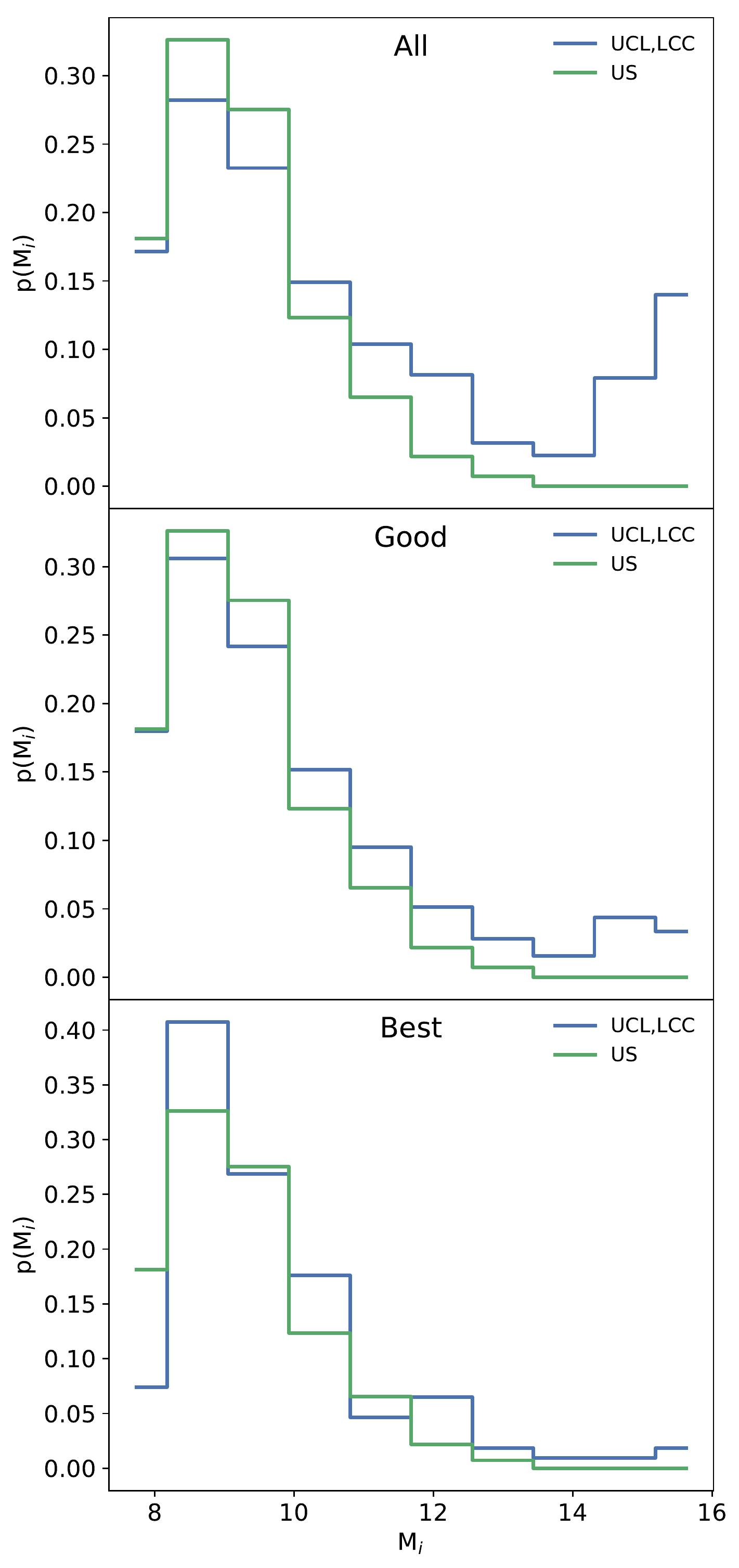}
    \caption{Comparison between the luminosity function measured using
      our UCL and LCC data and the luminosity function using
      \citet{Ardila2000} observations of Upper Sco, normalized so that
      each distribution has the same total number of stars in the
      magnitude range available to \citet{Ardila2000} (our observation
      range covers slightly brighter to significantly fainter
      objects), using the \citet{Jordi2006} offset of $0.329+0.29(R-I)$ to 
      convert \citet{Ardila2000}'s $I$ to our SDSS calibrated $i$ magnitudes.
      The $x$ axis displays the absolute magnitude in $i$,
      calculated assuming distances of 145 pc, 142 pc, and 118 pc for
      Upper Sco, UCL, and LCC, respectively. The y axis is the
      fraction of the total sources in the group (or in this case
      subset of UCL, LCC, or Upper Sco) in a given luminosity bin. We
      see that all three sets of proper motion cuts described in
      Appendix \ref{scoii-select-members} show excellent agreement with
      the Upper Sco luminosity function, although they do deviate from
      each other for sources fainter than 14 mag. We expect that this
      is due to interlopers in the top panel with no or poorly
      constrained proper motions.}\label{fig:lf_i_compare}
\end{figure}

This does not mean that the low mass LF is universal, as it can change
dramatically in time. \citet{Allen2003} investigated how the LF
changes in clusters as they age, simulating a cluster pulled from a
\citet{Kroupa2002} IMF and allowing the low mass stars and substellar
objects to evolve using the \citet{Burrows2001} models. As brown
dwarfs cease deuterium burning and begin to rapidly cool, their
luminosities decrease and distinct peaks in the group's luminosity
function can be seen as brown dwarfs transition from star-like
luminosities to planet-like luminosities (see \citealt{Allen2003} for
a description of the evolution of cluster luminosity up to 1
Gyr). \citet{Allen2003} compared observations of several young
clusters and associations to simulations which exhibit agreement with
a Kroupa-like power-law, with a low mass $\Gamma=0.5$. Rather than run
a new set of stellar evolutionary models, we use the
\citet{Ardila2000} observations of Upper Sco modeled by
\citet{Allen2003} as a tool for comparison: due to the similar ages of
Upper Sco, UCL, and LCC, as well as their origin in the same molecular
cloud, there should not be a significant difference between their
LF's. The main source of error in our LF are uncertainties in the
distance to the objects, where a difference between 100 pc and 200 pc
gives a systematic error of $\sim$1.5 mag, while the mean uncertainty
in apparent magnitude is $\sim$0.02 mag. To convert the
\citet{Ardila2000} $I$ to SDSS $i$ we use the transformation $i-I=(0.247 ± 0.003)*(R-I)$
from \citet{Jordi2006} (see Figure
\ref{fig:lf_i_compare}).

\begin{figure}\centering
    \includegraphics[width=0.3\textwidth]{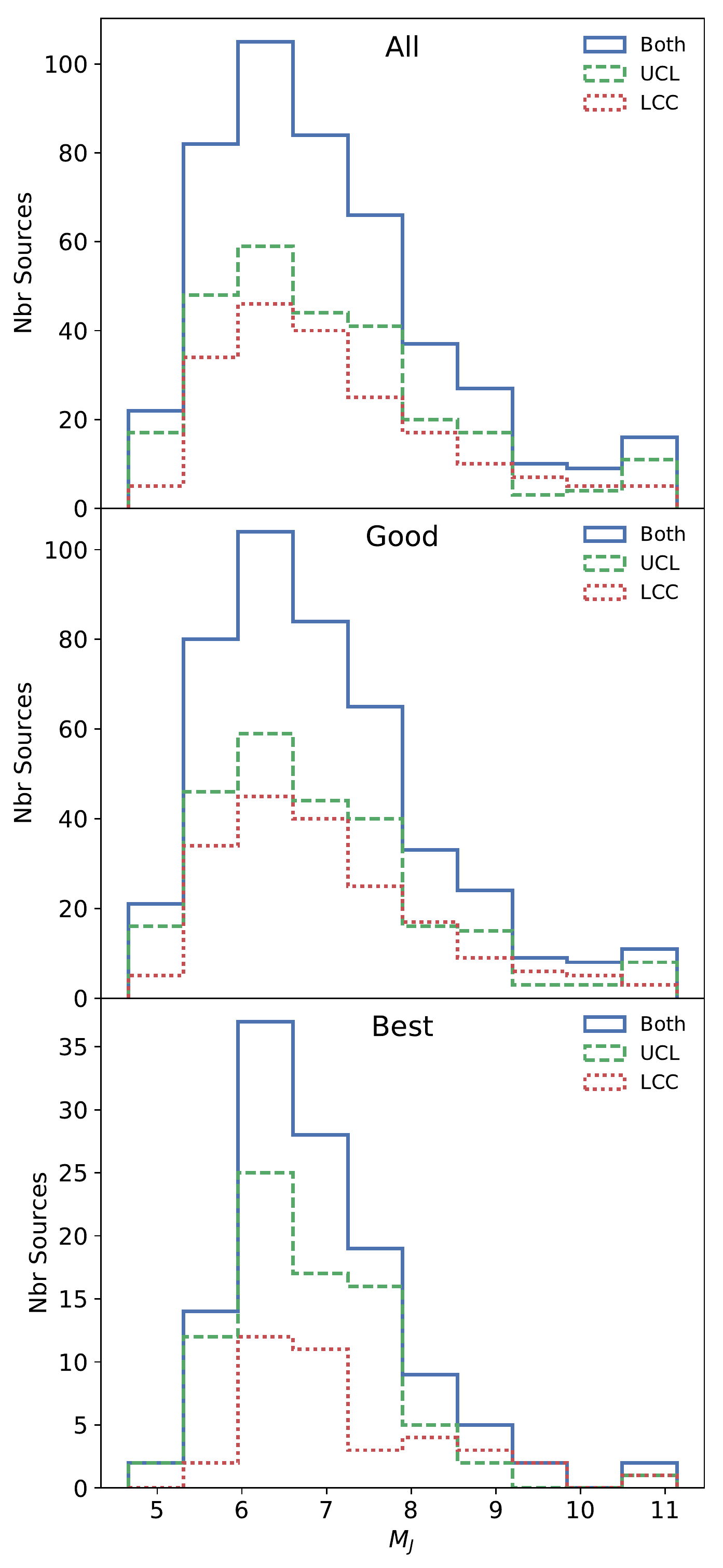}
    \caption{Comparison of the M$_{J}$ luminosity functions for LCC,
      UCL, and their combined luminosity function. The top plot shows
      all of the candidates objects, the middle combines the
      \emph{good} and \emph{best} sources, and the bottom plot shows
      only the \emph{best} sources.}\label{fig:imf_J}
\end{figure}

We also compare the $M_J$ luminosity function from UCL, LCC, and their
combined LF in Figure \ref{fig:imf_J}. Here we see the importance in
having a large number of sources to create the low mass LF in a
meaningful way, as the $M_J$ LF for LCC sources observed in two epochs
only has 38 \emph{best} sources and is noticeably different from the
UCL and combined LFs. Once a reasonable number of sources is included
in the sample (for example the 80 \emph{best} UCL sources) the LF has
the same approximate shape regardless of the number of sources added
or cuts used. This shows that while individual fields might show
slight variations, the underlying mass function inferred from the LF
is likely to be the same in all of Sco-Cen (Upper Sco, UCL and LCC).

\subsection{Mass Functions in UCL and LCC}\label{scoii-imf}

\begin{figure}
    \includegraphics[width=.3\textwidth]{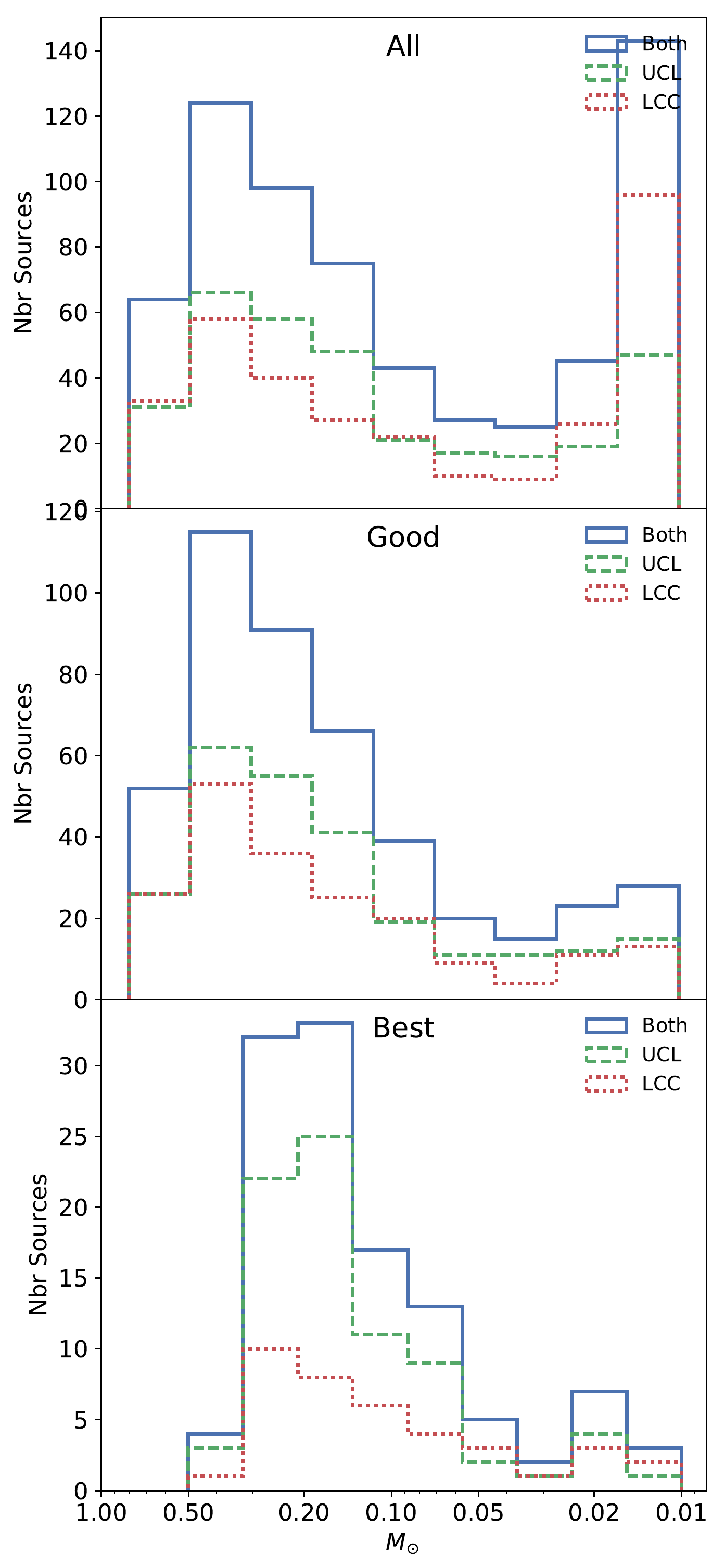}
    \caption{We compare the mass functions in UCL, LCC, and a
      combination of the two using the proper motion cuts. The
      \emph{best} sample contains enough sources to likely be the most
      representative of the actual IMF and future work will
      investigate the anomalous excess in the \emph{good} and ``all''
      plots for $m<20 M_\textrm{Jup}$.}\label{fig:imf_mass}
\end{figure}

Calculation of the mass function (MF) from a set of photometric observations is
more prone to error/uncertainty as it cannot be calculated without the aid of evolutionary models.
This is further complicated by known errors with most stellar evolution codes, which tend
to predict younger ages for lower mass objects and older ages for
higher mass stars \citep{Bell2012, Pecaut2012, Kraus2015, David2016}.
\citet{Rizzuto2016} analyzed a population of G,K, and M-type binary systems in Upper Scorpius
and estimated that while the G-type systems were consistent with
$\sim$11 Myr ages, M dwarfs appeared to be much younger ($\sim$7 Myr),
corresponding to an effective temperature over prediction of 100-300 K.
They also note that while the majority of B-type stars in Upper Scorpius
were consistent with an age of $\sim$11 Myr, a few B-type stars show
evidence of being 6-9 Myr younger.

\citet{Pecaut2016} analyzed the ages
of 657 F, G, K, and M-type members in all three Sco-Cen subgroups. Instead
of predicting the ages using a set of model isochrones, they use the
observed luminosities of the stars to calculate an empirical
isochrone. They assume that each subgroup has some mean age with an
intrinsic age spread around the mean, corresponding to different star-forming events.
Younger stars are more luminous, so for a given
spectral type they equate the mean luminosity with the mean age and
interpret a spread in luminosities as a spread in ages. This gave them
an empirical isochrone, which they used to generate a chronological
plot of the age distribution for the entire Sco-Cen association. As
predicted by \citet{Rizzuto2016}, they show that lower mass objects
appear to have their ages underestimated by stellar and substellar
models.

If the results of \citet{Rizzuto2016} and \citet{Pecaut2016} are
correct, the locations of our sources in color magnitude diagrams
should make them appear younger (as a whole) than the mean ages of UCL
and LCC. Looking back at Figure \ref{fig:lcc_ucl_candidates} we see
that indeed they do, where very few objects apear older than 15 Myr,
with a mean inferred isochronal age between 2-10 Myr and fainter
sources appearing much younger than even the 1 Myr isochrone(not shown in
Figure \ref{fig:lcc_ucl_candidates})).
\citet{Feiden2016} investigated the effects of magnetic fields on 
convection for a population of A-M stars in Upper Scorpius.
He showed that simulations introducing a magnetic perturbation,
with a peak magnetic energy density much less than the
thermal energy density, inhibits convection in low mass stars, causing
them to appear more luminous. Due to shrinking convection zones as the
mass of a star increases, this effect decreases and all but vanishes
for stars with masses above $\sim1.2$ $M_{\odot}$.

Because Feiden's models have a lower mass limit of 85 M$_\textrm{Jup}$, there are no
models covering our observed mass range that accurately match observations.
For this reason one has to be careful how to interpret the results of our MCMC
predictions of mass, age, and distance and  respect the various caveats inherent within.
The reason for calculating them at all is that the magnitudes in the
BT-Settl atmospheric models \citep{Allard2011} are given at the stellar surface,
requiring an isochrone (or set of isochrones) to calibrate the radius at each
T$_\textrm{eff}$ to calculate an absolute magnitude.
One might argue that fitting distance, age, A$_V$, mass, and $\log (g)$ is
overkill just to get an effective temperature, but spectral results to be
published in \citet{Moolekamp2018} following up on a few dozen
objects show that our predictions of spectral class are accurate to within
$\pm \frac{1}{2}$ a subclass (for most sources).
Examination of Figure \ref{fig:lcc_ucl_candidates} also shows that predicted spectral classes
for most sources transition smoothly from hotter to cooler sources for objects with
measured proper motions and $JHK_s$ photometry.
Unfortunately, slight changes in T$_\textrm{eff}$, distance, mass, and age, create
degeneracies in the photometry, making it much more difficult to calculate them
with any degree of certainty.

Even if we were to have a set of models that could reasonably predict the
mass of our sources using photometry alone, higher order corrections due to
binary fraction and dynamical evolution can have a noticeable impact on the IMF,
making a much more thorough study necessary to properly model the IMF in UCL and LCC.
Instead we present preliminary results from our BHAC2015 \citep{Baraffe2015} model comparisons
in Figure \ref{fig:imf_mass} to qualitatively display the
approximate MF using the \emph{best}, \emph{good}, and entire
candidate lists (see Appendix \ref{scoii-pm-cuts}).
Future work that includes spectroscopic observations
and a mass-luminosity relation derived from spectroscopic binaries in
Upper Scorpius will allow us to provide a more thorough estimate \citep{Moolekamp2018}.

\section{IR Excess in UCL and LCC}\label{scoii-disks}

\begin{figure}\centering
    \includegraphics[width=.4\textwidth]{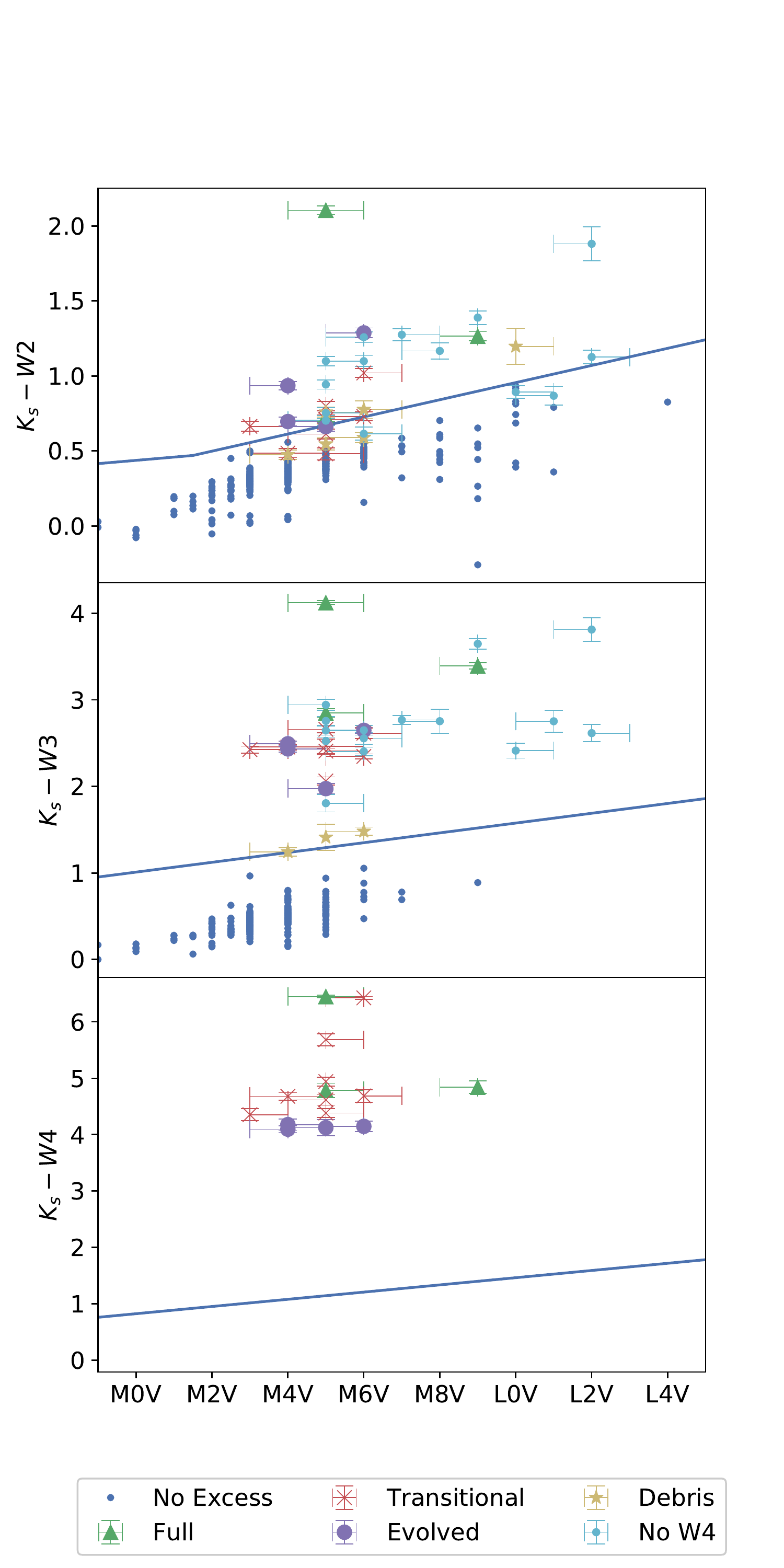}
    \caption{Color-color diagrams for 2MASS $K_S$ and AllWISE $W1$, $W2$, $W3$, and
      $W4$ compared with estimated T$_\textrm{eff}$,
      compared to \citet{Baraffe2015} isochrones using the BT-Settl
      \citep{Allard2011} atmospheric models (black solid line) and 2$\sigma$ cutoff (blue dashed line).
      Candidates have been categorized based on the observational criteria
      in \citet{Luhman2012}, with small dots representing sources that do not
      show excess indicative of circum-primary disks.
      Sources with error bars show an excess in all three WISE colors and are
      likely to have circum-primary disks (see Table \ref{tab:wise}). }\label{fig:wise}
\end{figure}

\begin{figure}\centering
    \includegraphics[width=.4\textwidth]{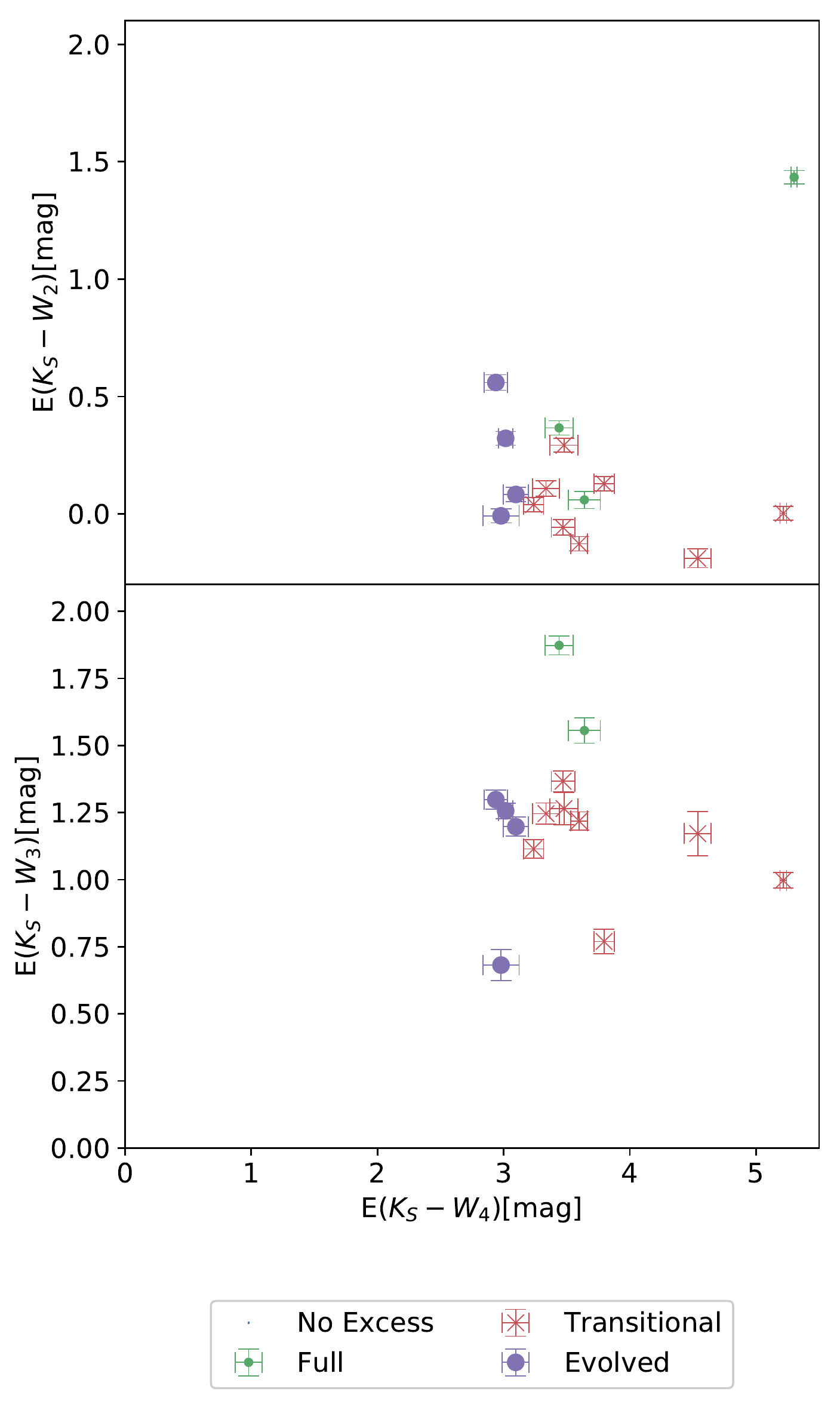}
    \caption{Color excess and estimated disk types for all objects with excess in all WISE bands}\label{fig:wise-colors}
\end{figure}

As mentioned in the introduction, 10-20 Myr is an important timeframe in the
lifetime of circum-primary disks, where stars (and presumably substellar objects)
can have disks in a wide range of evolutionary stages.
For objects with optical spectra, the detection of disks can be
performed by calculating the equivalent width of the H$\alpha$ feature \citep{Barrado2003}
and using it to classify a star as a Classical or Weak T-Tauri Star (CTTS or WTTS).
It is also possible to identify disks using IR photometry,
where \citet{Espaillat2012} used \emph{Spitzer IRAC} \citep{Fazio2004} and \emph{WISE}
photometry to distinguish between full, transitional, evolved, debris, and no circumstellar disks.
Photometric classification of disks uses IR photometry compared to K-band photometry, which is short enough to resemble photospheric flux and long enough to suffer from limited extinction, to estimate the excess color in a source compared to photospheric colors.

\begin{table*}
    \captionsetup{justification=centering}
    \caption{Infrared excess and disk fractions for K-type and M-type member of UCL and LCC. Using the criteria described in Section \ref{scoii-disks}, we show the fraction of K-type members from \citet{Pecaut2016} and compare them to the M-type members in this paper. Note that the \citet{Pecaut2016} sources have been spectroscopically confirmed with EW(H$\alpha$ measurements) while the M-type stars from this survey are candidate members with estimated spectral types and our estimated disk fractions should be viewed as lower bounds, as many of our sources do not have reliable $W3$ and $W4$ photometry (due to the small number of sources with reliable $W4$ photometry we do not include the percentage of sources with excess in $W4$ to avoid confusion).}\label{tab:disks}
    \centering
    \begin{tabular}{|l|r|r|r|r|}
        \hline
        \multirow{2}{*}{Band/Disk Type} & \multicolumn{2}{|c|}{UCL} & \multicolumn{2}{|c|}{LCC} \\
        \cline{2-5}
        & K-type & M-type & K-type & M-type \\
        \hline
        W2 & 11/157 (7.0\%) &  14/246 (5.7\%) & 5/119 (4.2\%) & 12/175 (6.9\%) \\
        W3 & 9/157 (5.7\%) & 16/139 (11.5\%) & 4/118 (3.4\%) & 11/129 (13.2\%) \\
        W4 & 44/72 & 8/8 & 29/86 & 7/7\\
        Full & 8/157 (5.1\%) & 3/267 (1.1\%) & 4/118 (3.4\%) & 0/199 (0.0\%) \\
        Transitional & 0/157 (0.0\%) & 4/267 (1.5\%) & 0/118 (0.0\%) & 4/199 (2.0\%) \\
        Evolved & 1/157 (0.6\%) & 1/267 (0.4\%) & 0/118 (0.0\%) & 3/199 (1.5\%) \\
        Debris & 24/157 (15.2\%) & 5/267 (1.9\%) & 14/118 (11.9\%) & 1/199 (0.5\%) \\
        No W4 & -- & 6/267 (2.2\%) & -- & 9/199 (4.5\%) \\
        \hline
    \end{tabular}
\end{table*}

\begin{table*}
\caption{AllWISE photometry for sources with 2$\sigma$ excess in $K_s-W1$ $K_s-W2$, $K_s-W3$, $K_s-W4$ colors. Sources marked with a $\dagger$ show an excess in that color.\label{tab:wise}}
\begin{tabular}{ccccccc}
\hline \hline
id & AllWISE id & Spectral Template & $K_S-W2$ & $K_S-W3$ & $K_S-W4$ & Disk Type \\
 &  &  & (mag) & (mag) & (mag) &  \\
\hline
Moolekamp 2 & J121551.25-493734.8 & M6 & $0.36\pm 0.04$ & $2.30\pm 0.07$$\dagger$ & -- & No W4 \\
Moolekamp 13 & J121610.32-521919.7 & M6 & $0.56\pm 0.03$$\dagger$ & $1.92\pm 0.04$$\dagger$ & $3.42\pm 0.09$$\dagger$ & Evolved \\
Moolekamp 17 & J122351.30-522413.4 & M7 & $0.61\pm 0.04$$\dagger$ & $2.11\pm 0.05$$\dagger$ & -- & No W4 \\
Moolekamp 45 & J121925.27-562049.9 & L2 & $0.76\pm 0.11$$\dagger$ & $2.69\pm 0.14$$\dagger$ & -- & No W4 \\
Moolekamp 49 & J122242.29-563611.5 & M5 & $0.41\pm 0.03$$\dagger$ & $2.11\pm 0.03$$\dagger$ & $4.09\pm 0.08$$\dagger$ & Transitional \\
Moolekamp 73 & J152649.42-355829.2 & M5 & $0.48\pm 0.03$$\dagger$ & $2.72\pm 0.06$$\dagger$ & -- & No W4 \\
Moolekamp 76 & J152700.55-360113.3 & M5 & $0.21\pm 0.04$ & $1.08\pm 0.15$$\dagger$ & -- & Debris \\
Moolekamp 91 & J152842.76-362618.0 & M5 & $0.27\pm 0.04$ & $2.26\pm 0.08$$\dagger$ & $5.48\pm 0.11$$\dagger$ & Transitional \\
Moolekamp 105 & J152907.03-365551.8 & M5 & $0.46\pm 0.03$$\dagger$ & -- & -- & Debris \\
Moolekamp 121 & J153516.17-393746.3 & L0 & $0.70\pm 0.12$$\dagger$ & -- & -- & Debris \\
Moolekamp 153 & J154609.72-344827.9 & M4 & $0.22\pm 0.03$ & $0.99\pm 0.05$$\dagger$ & -- & Debris \\
Moolekamp 167 & J154745.85-351319.7 & M5 & $0.38\pm 0.03$ & $1.70\pm 0.06$$\dagger$ & $3.85\pm 0.14$$\dagger$ & Evolved \\
Moolekamp 169 & J154756.93-351434.9 & M5 & $0.83\pm 0.03$$\dagger$ & $2.85\pm 0.02$$\dagger$ & $5.17\pm 0.02$$\dagger$ & Full \\
Moolekamp 180 & J154816.78-352321.3 & M9 & $0.73\pm 0.03$$\dagger$ & $2.86\pm 0.04$$\dagger$ & $4.31\pm 0.11$$\dagger$ & Full \\
Moolekamp 181 & J154826.33-352544.3 & M9 & $0.80\pm 0.05$$\dagger$ & $3.06\pm 0.06$$\dagger$ & -- & No W4 \\
Moolekamp 185 & J154525.41-353431.5 & M5 & $0.39\pm 0.03$$\dagger$ & $2.45\pm 0.05$$\dagger$ & -- & No W4 \\
Moolekamp 200 & J154806.22-351548.5 & M6 & $0.41\pm 0.03$$\dagger$ & $2.03\pm 0.03$$\dagger$ & $6.11\pm 0.03$$\dagger$ & Transitional \\
Moolekamp 224 & J125232.27-563053.1 & L0 & $0.43\pm 0.04$ & $1.96\pm 0.09$$\dagger$ & -- & No W4 \\
Moolekamp 225 & J124541.67-564312.0 & M6 & $0.57\pm 0.03$$\dagger$ & $2.16\pm 0.06$$\dagger$ & $4.24\pm 0.11$$\dagger$ & Transitional \\
Moolekamp 248 & J125423.62-531111.2 & M4 & $0.26\pm 0.03$ & $2.23\pm 0.03$$\dagger$ & $4.45\pm 0.07$$\dagger$ & Transitional \\
Moolekamp 337 & J123410.32-513042.3 & L1 & $0.64\pm 0.06$ & $2.52\pm 0.13$$\dagger$ & -- & No W4 \\
Moolekamp 347 & J122651.99-523618.1 & L2 & $0.78\pm 0.05$$\dagger$ & $2.27\pm 0.10$$\dagger$ & -- & No W4 \\
Moolekamp 366 & J123642.32-503613.2 & M6 & $0.57\pm 0.04$$\dagger$ & $1.96\pm 0.06$$\dagger$ & -- & No W4 \\
Moolekamp 369 & J123938.39-504240.4 & M6 & $0.53\pm 0.04$$\dagger$ & $1.84\pm 0.05$$\dagger$ & -- & No W4 \\
Moolekamp 392 & J120808.89-502655.9 & M4 & $0.43\pm 0.03$$\dagger$ & $2.16\pm 0.04$$\dagger$ & $3.91\pm 0.10$$\dagger$ & Evolved \\
Moolekamp 396 & J120624.51-505415.3 & M4 & $0.46\pm 0.03$$\dagger$ & $2.02\pm 0.03$$\dagger$ & $3.62\pm 0.06$$\dagger$ & Evolved \\
Moolekamp 414 & J122556.71-472154.7 & M6 & $0.31\pm 0.03$ & $1.20\pm 0.05$$\dagger$ & -- & Debris \\
Moolekamp 449 & J130515.48-502438.9 & M5 & $0.57\pm 0.03$$\dagger$ & $2.11\pm 0.05$$\dagger$ & -- & No W4 \\
Moolekamp 466 & J131107.55-445553.1 & M5 & $0.43\pm 0.03$$\dagger$ & $1.70\pm 0.05$$\dagger$ & $4.57\pm 0.08$$\dagger$ & Transitional \\
Moolekamp 524 & J150703.47-440859.3 & M5 & $0.37\pm 0.04$$\dagger$ & $1.43\pm 0.10$$\dagger$ & -- & No W4 \\
Moolekamp 590 & J143000.21-453045.3 & M8 & $0.55\pm 0.05$$\dagger$ & $2.14\pm 0.14$$\dagger$ & -- & No W4 \\
Moolekamp 597 & J142844.81-455718.0 & M6 & $0.52\pm 0.06$$\dagger$ & -- & -- & Debris \\
Moolekamp 602 & J142429.95-462142.1 & M3 & $0.36\pm 0.03$$\dagger$ & $2.12\pm 0.04$$\dagger$ & $4.05\pm 0.11$$\dagger$ & Transitional \\
Moolekamp 619 & J140445.33-432348.4 & M5 & $0.52\pm 0.03$$\dagger$ & $2.10\pm 0.05$$\dagger$ & -- & No W4 \\
Moolekamp 630 & J140020.08-440317.6 & M5 & $0.42\pm 0.04$$\dagger$ & $2.54\pm 0.05$$\dagger$ & $4.48\pm 0.13$$\dagger$ & Full \\
Moolekamp 652 & J140635.06-392640.9 & M5 & $0.34\pm 0.03$ & $2.39\pm 0.04$$\dagger$ & $4.34\pm 0.09$$\dagger$ & Transitional \\
\hline
\end{tabular}
\end{table*}

The X-ray and kinematic search of Sco-Cen performed by \citet{Mamajek2002}
used 2MASS $K_s$-band IR excess and H$\alpha$ emission to search for accretion disks
around K-type members of UCL and LCC, identifying only a single source out of
110 stellar candidates that displayed evidence of an optically thick accretion disk.
More recently \citet{Pecaut2016} used the \citet{Espaillat2012} classification scheme,
combined with the \citet{Luhman2012} observational criteria to estimate the disk fraction
for K-type stars in UCL and LCC (they also had a few early M stars in their sample but not enough to provide a reasonable estimate of the disk fraction), where they identify 12 out of 275 stars K stars with a
full disk, none with a transitional disk, and a single evolved disk.
If estimates that lower mass stars can possess longer lived disks is correct, we should expect
to find more disks around the M and L dwarfs in our sample.

We perform an analysis similar to \citet{Pecaut2016}, using the \citet{Espaillat2012} classification
criteria and the \citet{Luhman2012} observational criteria.
$K_S$ photometry is taken from the \emph{2MASS} catalog and compared to \emph{AllWISE} $W2$, $W3$, and $W4$ (in this mass regime excess in $K_S-W1$ due to the presence of a disk is negligible) for each source.
We reject all sources with photometric errors $>0.2$ mag and visually inspect all remaining sources for nearby neighbors and other contaminants.
Of the 466 good or best sources: 421 have reliable $W2$ photometry, 268 have $W3$, and only 15 have $W4$ photometry with errors $<0.2$ and no local contaminants.
We characterize IR excess in our sample using the same criteria as \citet{Luhman2012} used for Upper Sco, where an excess in $K_S-W2$ is defined by the lines connecting (B0, 0.19), (K0, 0.22), (M1.5, 0.47), and (M8.5, 0.87), $K_S-W3$ excess is defined by (B0, 0.18), (G8, 0.33), and (M9, 1.52), and $K_S-W4$ excess is bounded by (B0, 0.11), (K6, 0.57), and (M9, 1.4) (because some sources in our sample have estimated spectral types later than M9, we interpolate the last line segment for each color to L5).
This is slightly different than the criteria used in \citet{Pecaut2016}, so we also re-analyze their sample of K stars in UCL and LCC for a more accurate comparison with our M dwarf candidates.
Using the same excess boundaries as \citet{Luhman2012} also allows us to use the same criteria to classify disks, where we classify a disk as
``full'' if E($K_S-W3$) > 1.5 and E($K_S-W4$) > 3.2,
``transitional'' if E($K_S-W3$) < 1.5 and E($K_S-W4$) > 3.2,
``evolved'' if E($K_2-W3$) > 0.5 and E($K_S-W4$) < 3.2,
and ``debris'' if E($K_S-W3$) < 0.5 and E($K_S-W4$) < 3.2.
Most of our sources do not have reliable $W4$ photometry, so we add an additional classification for our candidates with $K_S-W3$ > 0.5 but no reliable $W4$ photometry as ``no $W4$''.
Table \ref{tab:wise} shows the $K_s-Wn$ colors for candidates showing an IR excess in one or more of the bands, with a $\dagger$ marking colors that show an excess from estimated photospheric colors, and the estimated disk type of each source using the criteria stated above.

Table \ref{tab:disks} shows a comparison of IR excess and disk fractions from our survey of M dwarfs in UCL and LCC compared to
\citet{Pecaut2016} observations of K-type stars in the same subgroups
(while the \citet{Pecaut2016} sample contains roughly the same number of objects, due to stellar populations
their sample is obtained over a much larger region of UCL and LCC).
We find comparable IR excess in $W2$ for K- and M-type sources and while we seem to find a higher fraction of sources with $W3$ and $W4$ excess, we are biased to find a higher percentage of objects with excess since lower mass objects are less likely to have reliable photometry in those bands without excess flux.
While we appear to find more transitional and evolved disks and fewer full and debris disks, excess calculations are highly dependent on spectral type and proper spectral classifications might correct our results.
We expect to revisit this issue in more detail in a future paper once we have spectrally classified a large portion of our candidates.

\section{Mass Segregation in $\eta$ Cha} \label{scoii-etacha}

In addition to the main subgroups of Sco-Cen: Upper Sco, UCL, and LCC,
there are a number of nearby clusters kinematically linked to the
Sco-Cen star-forming complex. Among these is the 11$\pm$3 Myr old
$\eta$ Cha cluster \citep{Bell2015}, first discovered
by \citet{Mamajek1999} using X-ray detections from ROSAT. At a
distance of $\sim$95 pc, $\eta$ Cha has higher proper motions than the
main Sco-Cen subgroups and has negligible reddening (Av$\sim$0 mag),
making it easier to kinematically and photometrically select
members. It is also a relatively compact cluster, almost entirely
contained in a single DECam image (see Figure \ref{fig:all_fields}),
with only 18 known members \citep{Becker2013}.

\begin{table*}
\caption{Members detected in $\eta$ Cha by the SCOCENSUS pipeline. The
  \emph{Spectral Type} is the spectral type as given in
  \citet{Luhman2004}, while the \emph{Estimated Sp Type Range} is the
  predicted range of spectral types using MCMC and the \emph{Spectral
    Template} is the most likely spectral
  type.\label{tab:scoii-etacha}}
\begin{tabular}{cccccc}
\hline \hline
SCOCENSUS & 2MASS & RECX & SpT & Estimated SpT Range & Spec. Template \\
\hline
100-9452  & J08440914-7833457 & 16 & M5.75 & M6-M6 & m6 \\
100-24176 & J08385150-7916136 & 17 & M5.25 & M5-M6 & m5 \\
100-40810 & J08413030-7853064 & 14 & M4.75 & M4-M5 & m5 \\
\hline
\end{tabular}
\end{table*}

\begin{figure}
    \includegraphics[width=.48\textwidth]{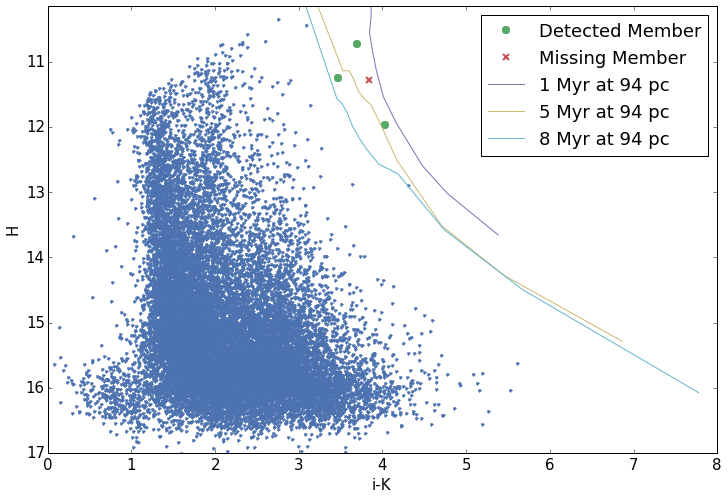}
    \caption{Color magnitude diagram for $\eta$ Cha field (F100). The
      green circles mark the $\eta$ Cha members that we detected and
      the red x is an M dwarf that was in between the CCD gaps in our
      2015 images (the point is plotted using the magnitudes from \citet{Luhman2004}).
      The predicted age range of $\eta$ Cha is $\sim$6-8
      Myr but based on the discussion in Section \ref{scoii-imf}, it
      seems likely that $\eta$ Cha is closer to the 11 Myr age
      predicted by \citet{Bell2015}, since the \citet{Baraffe2015}
      models tend to underpredict the age of low mass stars.}
    \label{fig:etacha-ikh}
\end{figure}

Previous surveys of the region performed by \citet{Luhman2004,
  Song2004, Lyo2006} have searched for substellar objects in $\eta$
Cha but no objects $\lesssim$0.1 M${}_{\odot}$ were found within
1.5${}^{\circ}$. \citet{Murphy2010} extended the search in a wider
survey with a radius of 5.5${}^{\circ}$, finding 4 ``probable'' and 3
``possible'' members with estimated masses between 0.08 and 0.3
M${}_{\odot}$. \citet{Becker2013} noted these results and attempted to
reverse engineer the dynamical evolution of $\eta$ Cha assuming a
universal IMF and the known positions of current members. By running a
number of N body simulations using several sets of initial conditions,
they concluded that it is very unlikely that $\eta$ Cha's current
configuration, with high mass stars concentrated toward the center and
low mass stars in a surrounding halo, occurred by dynamical processes
alone. Instead they conclude that the mass segregation in the region
is most likely primordial.

To investigate this claim we took a single set of DECam images
centered on $\eta$ Cha (field \emph{F100}). We ran the images through
the same pipeline used for our UCL and LCC fields and detected three M
dwarfs: \emph{2MASS J08440914-7833457}, \emph{2MASS
  J08385150-7916136}, and \emph{2MASS J08413030-7853064}, all of which
were detected by \citet{Luhman2004} (they also detect an additional fourth M-dwarf
that is not in our catalog, see Figure \ref{fig:etacha-ikh}). The
missing M dwarf fell between the gaps of our detector, as we did not
dither our images in order to stack individual CCD's. We did not
detect any new brown dwarfs down to magnitudes $i=21.2$, $z=20.3$,
$Y=18.6$ mag, including objects fainter than 2MASS and AllWISE, verifying
that there are not likely to be any brown dwarfs within a
1.1${}^{\circ}$ radius of $\eta$ Cha.

This field also served as a valuable crosscheck for our pipeline. All
three M dwarfs in our catalog were flagged as $\eta$ Cha candidates by
our pipeline and the spectral types of the objects fall within the
ranges of our predictions (see Table \ref{tab:scoii-etacha}). We
should note that these would be flagged as \emph{good} and not
\emph{best} by our pipeline, since none of the sources have DECam and
sky proper motions that are both consistent with Sco-Cen (two of the
sources have underestimated $\mu_{\alpha}$ and the other fell between
the detector gaps in 2013 and does not have a DECam proper
motion). The detection and proper spectral typing of all of the known
objects in our field of view serves as evidence that our photometry
and astrometry are well calibrated.

\section{Conclusion}\label{Conclusion}

We presented the results of a small scale $izY$ survey covering
$\sim$87 deg$^2$ in the UCL and LCC subgroups of the Scorpius
Centaurus OB association to search for low mass stars and substellar
objects. Our observations include a catalog with over 11 million
point sources and 466 candidate members of Sco-Cen, including an
estimated 80-100 brown dwarfs detected in LCC and UCL based on isochronal
models.

Our observations add to the growing body of evidence suggesting that
models of PMS stars and young substellar objects are missing key
physical processes that underestimate their luminosity in the near IR,
causing them to predict much younger ages for lower mass
objects. Nevertheless we show that the luminosity function in UCL
and LCC is nearly identical to observations made by \citet{Ardila2000}
in Upper Sco and that the IMF is consistent with observations of other
young clusters.

We see IR excess indicative of circum-primary disks, with similar disk fractions to higher mass K-type stars presented in \citet{Pecaut2016}.
There is some discrepancy in the exact classification, where we find more transitional and evolved disks and fewer full and debris disk.
Spectral classification of our objects will reveal if this is a real effect or a result of small differences between our estimated spectral templates and the objects true spectral types.

While the initial results of this survey are encouraging, there is
still much work to be done. Follow-up photometry is needed for objects
below the completion limit of GAIA, 2MASS, and AllWISE to calculate proper
motions for candidate L dwarfs ($m\lesssim$10-15
$M_\textrm{Jup}$). Using longer exposures in $zY$ bands it should also
be possible to detect later L dwarfs if they exist in isolation,
however in the absence of faint all-sky catalogs it will require
multiple epoch photometry to generate reasonable proper motions (for example \emph{LSST}).
Follow-up photometry in $JHK_s$ will also allow the removal of giant
interlopers in magnitude regions too faint for 2MASS detections.
This will provide valuable constraints to our photometrically selected L
dwarf candidates to allow us to probe the substellar IMF in a
currently unexplored mass regime. Spectroscopic followup of our
candidates has already begun and will help us understand any biases in
our candidate selection and estimations of spectral type and mass,
which can be used to derive a more precise model of the IMF in
Sco-Cen.

\section*{Acknowledgements}

FM and EEM acknowledge support from NSF award AST-1313029.

This research was carried out at the Jet Propulsion Laboratory, California Institute of Technology, under a contract with the National Aeronautics and Space Administration. EEM acknowledges support from the Jet Propulsion Laboratory Exoplanetary Science Initiative and the NASA NExSS Program.

KL was supported by grant AST-1208239 from the NSF.

We thank the staff of CTIO and the Blanco 4-m telescope in particular
for their help and hospitality during the 2013 and 2015 observing
runs.

We thank Frank Valdes and Robert Gruendl for their help understanding the
DECam community pipeline and calibration of photometry and astrometry.

The Center for Exoplanets and Habitable Worlds is supported by the
Pennsylvania State University, the Eberly College of Science, and the
Pennsylvania Space Grant Consortium.



\clearpage
\bibliographystyle{mnras}
\bibliography{paper} 



\appendix

\section{Image Reduction Pipeline}\label{scocensus}
When this survey was started there was no reliable pipeline to reduce such a large volume of data that was also tunable for the precise measurements required to accurately calculate proper motions and PSF photometry in the galactic plane.
In the meantime the LSST software stack has reached a higher level of maturity, especially with the addition of the new \emph{scarlet} deblender \citep{Melchior2018}, and is likely to outperform the data reduction described in this appendix.
This appendix is still included to describe the reduction used in this paper and educate the reader curious about how to reduce survey data at scale.

The computational framework needed to conduct our survey consists of a
collection of python libraries. Many of these libraries were developed
by the the first author but extensive use was made of \emph{astropy}
\citep{Astropy2013} and its affiliated packages (including
contributions to those packages made by the author). Because many of
the challenges we faced will confront other groups both in the present
and future, we decided to make our software open source and more
general than necessary for our own use, but not as general as
possible. This was done to save time and because anyone performing
interesting research is likely to need slight modifications to the
codes anyway and will likely need to fork our code and use their own
version. The main framework and execution of the pipeline is handled
by the \emph{datapyp}
package \footnote{\url{https://github.com/fred3m/datapyp}}, written by
the first author to enable users to run large pipelines so that when
the inevitable crash occurs, users can easily modify the code and
restart the pipeline at any previous (or future) step. It is also
possible to run subsets of a pipeline, making it much easier to make
improvements to image reduction and analysis portions of the
pipeline. The vast majority of our pipeline is available as the
\emph{astropyp}
package \footnote{\url{https://github.com/fred3m/astropyp}}, which
contains the general high-level image reduction procedures that might
be useful to a wide variety of astronomers. Functions more specific to
our own research can be found in the \emph{scocensus} (SCOrpius
CENtaurus SUbstellar Survey) package, which includes all of the
scripts used to run our pipeline and select substellar objects from
our final catalog (this package is currently private because our data
is already public, but the software will be made public as well once
our survey is complete).

\subsection{Astrometry and Stacking}\label{scocensus-astrometry}

\begin{figure}
    \begin{center}
        \includegraphics[width=0.2\textwidth]{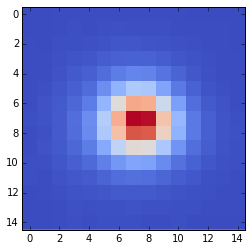}
        \includegraphics[width=0.2\textwidth]{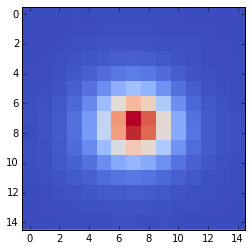}
    \end{center}
    \caption{The same source from a stack using \emph{SCAMP} and
      \emph{SWarp} (left) vs a \emph{astropyp} stack (right). By
      looking at a single source it is easier to see the improvement
      made to our PSF by stacking our images by using image
      coordinates.}
    \label{fig:stack2}
\end{figure}

The limiting factor in calculating sky positions of DECam sources is
the accuracy of the reference catalog used to generate the astrometric
solution. While DECam has better than 20 mas astrometric precision
\citep{Bernstein2015}, until the recent GAIA DR1 release \citep{GAIA2016},
the best all-sky catalog that covered Sco-Cen in
the magnitude range probed in exposures ranging from a few seconds to
several minutes was 2MASS \footnote{While some of the brighter sources
  appear in UCAC4, most of the UCAC4 errors in
  our magnitude range are larger than those by 2MASS and the overall
  astrometric solution is worse}, which has >100 mas uncertainty in RA
and DEC. Since the size of a DECam pixel is $\sim$262.7 mas/pixel, the
error is nearly half a pixel, causing stacked images to have
unnecessary noise (see Figure \ref{fig:stack2}).
Since our initial analysis, GAIA DR1 was published, however due to the
success of our astrometric solution we have not modified our pipeline
other than to use GAIA in the place of 2MASS as our reference catalog.

To improve the calculation of our proper motions we realized that it
would be necessary to create our own astrometric solution to align
exposures based on their image coordinates, not their world
coordinates (see Section \ref{scocensus-calibration}). This solution
is also useful for stacking our exposures as it allows us to create
stacks with much smaller positional errors.

The basic astrometric solution is very similar to the one implemented
in \emph{SCAMP}, using the transformations described in
\citet{Kaiser1999}:
\begin{align}
    x_r &= \sum_{l=0}^{l=N} \sum_{m=0}^{m=l} a_{lm} x^l y^m \\
    y_r &= \sum_{l=0}^{l=N} \sum_{m=0}^{m=l} b_{lm} y^l x^m
\end{align}
where $x_r,y_r$ are the reprojected positions of a source (which can
be any orthogonal 2D coordinate system), $x$ and $y$ are the source
positions in the original image, $a_{lm}$ and $b_{lm}$ are the
transformations polynomial coefficients, and $N$ is the order of the
polynomial. So the problem reduces to solving for $a_{lm}$ and
$b_{lm}$, which can be solved in a number of different ways. If we
were guaranteed that all of our source and reference positions were
accurate with well characterized errors, the best solution would be a
$\chi^2$ minimization of the transformation polynomials. This not only
depends on how well our pipeline chooses sources but also how high
their proper motion is. Since brighter stars are more likely to be
closer to us (and thus have higher proper motions), simply measuring
the positional errors is not sufficient to achieve the best fit (since
2MASS doesn't include proper motions and 90\% of the GAIA sources in our fields
lack proper motion measurements). We found that using Bayesian
analysis with MCMC (using the method described in \citealt{Hogg2010})
does a \emph{slightly} better job calculating the astrometric solution
(by a few mas at most), but since MCMC requires significantly more
time to run (remember we have to fit all 60 CCDs for each image), we
found that in the end a $\chi^2$ fit with a few user-defined cuts is
sufficient.

\begin{figure*}\centering
    \includegraphics[width=0.8\textwidth]{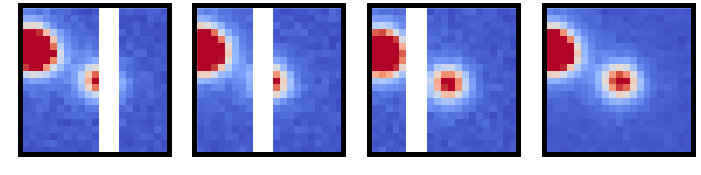}
    \caption{Elimination of bad pixels in the stack. The first three
      images from the left are single 200s exposures of a source that
      is bisected by a bleed trail on the CCD. The image on the right
      is taken from the stacks, which has no bad pixels since none of
      the images had overlapping data quality masks.}
    \label{fig:stack_mask}
\end{figure*}

Since the exposures used to create our stacks are only slightly
rotated and translated, a linear transformation is sufficient to
create the astrometric solution. This is not true later in our
pipeline when we compare images from different epochs and need a
higher order polynomial to calculate proper motions (see Section
\ref{scocensus-pmcal}). Once the solution has been derived the
pipeline creates a bivariate spline function to calculate flux as a
function of pixel coordinates in the original images. The bivariate
spline function is then used to map all of the pixels from the 3
images onto a common coordinate system (slightly larger than the
reference image so that all 3 images are contained in their entirety).

\begin{figure}
    \begin{center}
        \includegraphics[width=0.4\textwidth]{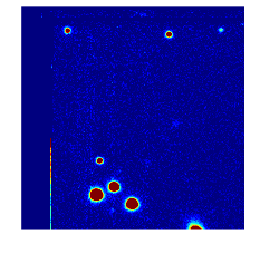}
        \includegraphics[width=0.4\textwidth]{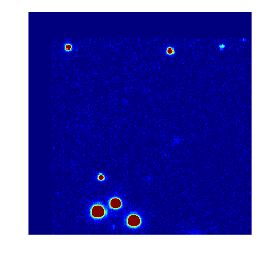}
    \end{center}
    \caption{Stacked images using \emph{SCAMP} and \emph{SWarp} (top)
      vs \emph{astropyp} (bottom). By combining images using image
      coordinates instead of the less accurate world coordinates
      (calibrated to GAIA) we are able to significantly improve the
      quality of our stacked images.}
    \label{fig:stack1}
\end{figure}

To minimize the effects of bad pixels, the data quality mask derived
by the community pipeline is used to mask the image arrays. The mean
of each unmasked pixel is used in the stack and any pixels that were
masked in all 3 images are masked in the final image and added to a
new data quality mask created for the stack. This allows us to save a
large number of sources that are partially obstructed by
non-overlapping bleed trails in all 3 images (see Figure
\ref{fig:stack_mask}). The final result is an image that has
noticeably less noise than the stacks we built using \emph{SCAMP} and
\emph{SWarp} (see Figures \ref{fig:stack2} and \ref{fig:stack1}).

When the final GAIA source catalog is released in 2022, the process
for reprojecting and stacking our images might be unnecessary, since
GAIA aims to have better than 10mas errors (which is better than the
errors on DECam image positions). It is even possible that GAIA DR1
may be a large enough improvement to render this procedure unnecessary,
or at the very least computationally
expensive for a minimal gain in accuracy (our stacks were built before the
GAIA DR1 release, so we have yet to test using it to create new stacks).
Otherwise this may be the
best solution to stack images and calculate proper motions (see Section
\ref{scocensus-calibration} for more on calculating proper motions
from an astrometric solution).

\subsubsection{Windowed Positions}

\begin{figure}\centering
    \includegraphics[width=0.4\textwidth]{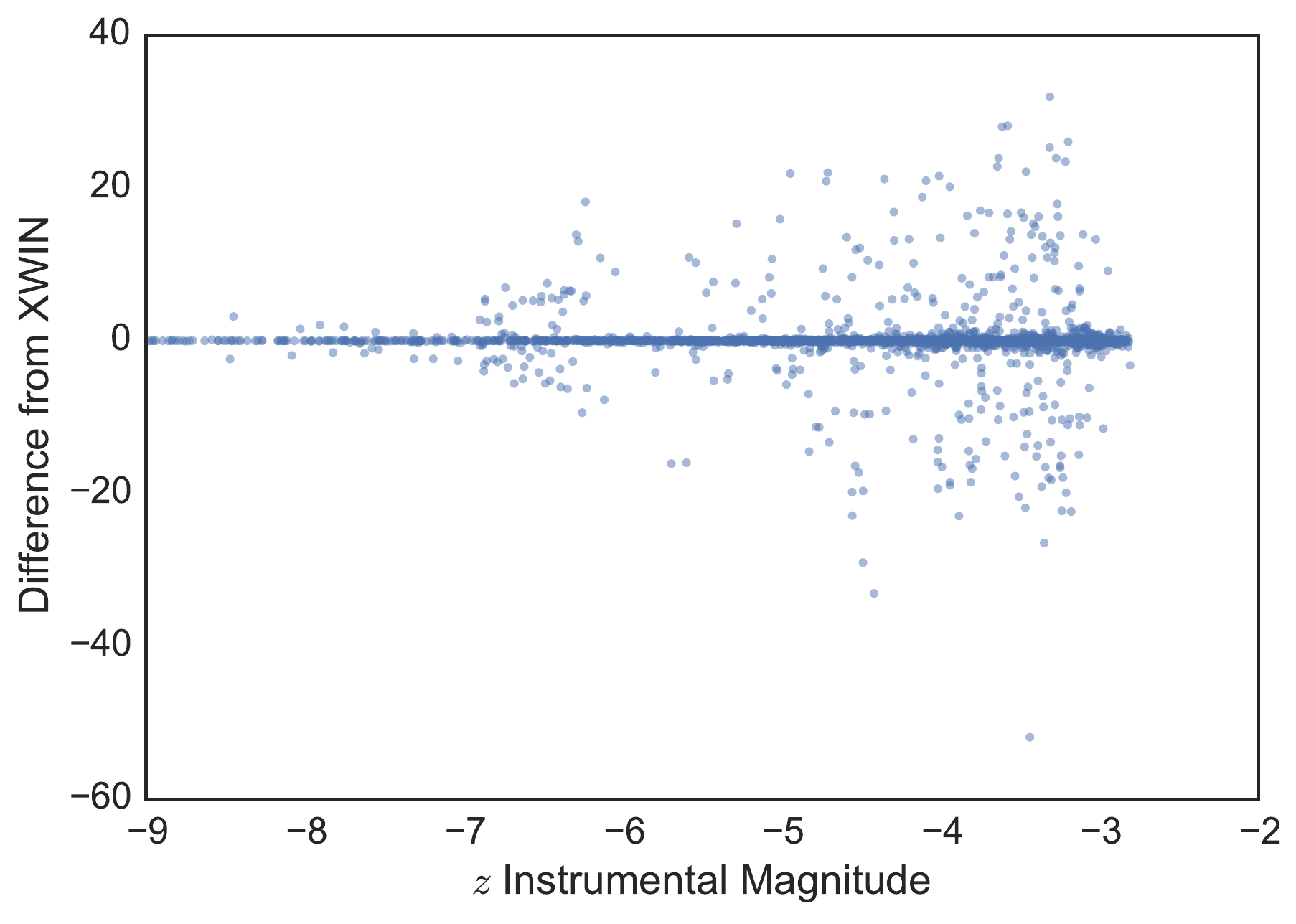}
    \includegraphics[width=0.4\textwidth]{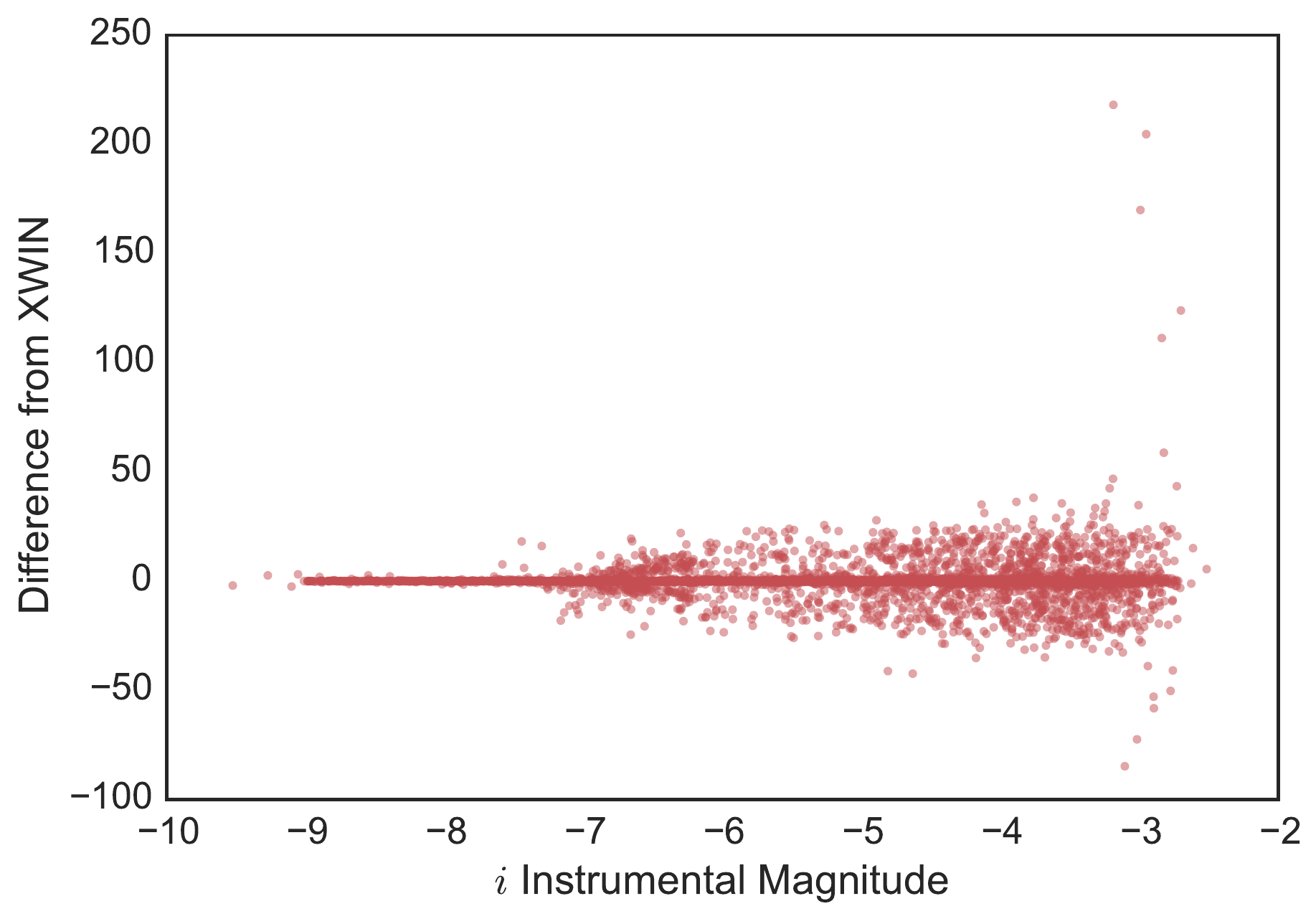}
    \caption{Difference in pixels between the x coordinate detected by
      \emph{SEP} and the windowed coordinate XWIN for a single CCD
      (similar differences are seen with the y and YWIN
      positions). The shorter (30s) z band exposure has 230 of 2074
      sources ($\sim$11\%) with a change in position of more than 1"
      while the longer (200s) i band exposure has 1233 of 4716 sources
      ($\sim$26\%) with a $>$1" coordinate change. Investigation of
      this effect appears to occur mostly in crowded fields or when a
      nearby neighbor is not detected by \emph{SEP}, which is why we
      see more frequent bad windowed positions in the longer
      exposures.}
    \label{fig:winpos}
\end{figure}

Another important feature we added to our code was an improvement to
the calculation of windowed positions by \emph{SEP}/\emph{SExtractor} (see \autoref{scocensus-photometry}). \emph{SExtractor}
windowed positions are improved source positions calculated using an
iterative procedure to find the weighted average of a sources
flux. Usually (for $\gtrsim$95\% of detected sources) this is a much more accurate position
but as you can see in Figure \ref{fig:winpos}, for a small percentage
of sources the windowed position can drift from one point source to a
completely different object. Both \emph{SExtractor} and \emph{SEP}
have flags to mark sources that fail to calculate windowed positions,
but in crowded fields we have still observed a small percentage of
sources that see their windowed position jump to a nearby brighter
source. In order to correct these positions we ignore any windowed
positions shifting by more than either an arcsecond or half the
distance to the nearest neighbor, whichever is less. Any source whose
windowed position is shifted by a larger amount uses its original
SExtracted position with a ``badwinpos'' flag set in the final catalog
to note that it has a less accurate position. When calculating the
mean position (see Section \ref{scocensus-calibration}), observations
with bad windowed positions are only used if the same source has a bad
windowed position in all of the images in the same epoch, otherwise
only the good windowed positions are used to calculate the catalog
position.

\subsection{Photometry} \label{scocensus-photometry}

Source detection and aperture photometry is performed on individual
CCDs using the \emph{SEP} package, a python library derived from
\emph{SExtractor}, one of the fastest codes to detect point
sources. The recent \emph{SExtractor} port to python gives researchers
more control over which functions are run, an easier interface to
modify parameters, a better understanding of what \emph{SExtractor} is
doing, and more control over the format and contents of the output
catalog.

\begin{figure}
    \begin{center}
        \includegraphics[width=0.2\textwidth]{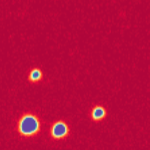}
        \includegraphics[width=0.2\textwidth]{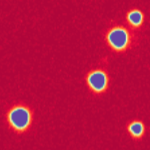}
    \end{center}
    \caption{Sample tiles from CCDs S29 (left) and N31 (right) from
      the same DECam image. The significant distortion in the PSF is
      obvious to the eye and is the reason we perform PSF photometry
      on each CCD independently.}
    \label{fig:focal_distort}
\end{figure}

Once the sources have been detected the next step is to automatically
select PSF stars. Beginning with the source catalog generated by
\emph{SEP}, we remove all sources that have a bad pixel anywhere in
their aperture or any other abberation flagged by the DECam community
pipeline data quality mask (for our stacks we only flag pixels that
were bad in the stack, see Section \ref{scocensus-astrometry} for more
on the bad pixel mask). To remove crowded stars we reject all sources
that have a neighbor within 3 times the detection aperture
radius. Although the CCDs near the edge of the DECam focal plane are
distorted (See Figure \ref{fig:focal_distort}), we found that
eliminating all sources with a best fit elliptical aperture ratio $>$
1.5 removes any galaxies or edge effects that may have been missed by
the data quality mask without rejecting good data. Finally we
eliminate all sources except those with a flux maximum $<$ 100 counts,
as DECam is known to be nonlinear in this regime
\citep{Bernstein2015}.

To create the PSF we use a similar procedure to the one implemented in
\emph{DAOPHOT}, loading a normalized subsampled image patch centered
on each PSF source. In most cases the exact center of the source is
not at the center of the patch, so all of the sources are recentered
on the subpixel with the maximum flux, giving a slightly cleaner PSF
than a PSF generated using the weighted average position. We then mask
all of the pixels outside the circular aperture used for the PSF and
store the pixelated PSF array as an \emph{astropy}
\emph{Fittable2DModel}.

One way to perform PSF photometry is to divide the image into groups
(or clusters) of sources, where sources with nearby neighbors are
lumped into the same group and PSF photometry is performed on the
entire group simultaneously (this is how DAOPHOT works). This can be
computationally expensive for very crowded fields, especially in
fields where nearly the entire CCD would be treated as a single group
(such as in the Galactic plane). It can also lead to instances, if the
variance is not accounted for sufficiently, where the brightest sources
in a group can dwarf the other sources and ruin the fit for fainter
stars.

Instead, we find all of the neighbors for each star that lie within 3
aperture radii ($\sim$10$''$). A patch from the data is extracted,
subsampled, and re-centered on the source to be fit, using the same
procedure used to generate the PSF. We then simultaneously fit the
target source (the one the patch is centered on) as well as it's
neighboring sources (that are only partially contained in the
patch). This allows for a much more accurate flux calculation and is
also much faster, as fitting an entire group simultaneously often
involves fitting a lot of pixels that are not part of one of the
sources (although this can be minimized by masking the pixels outside
the sources aperture radii). For very crowded images, where nearly all
of the sources would be contained in a single group (and thus
thousands of sources would need to be simultaneously fit), the group
would have to be subdivided into overlapping patches anyway.

\begin{figure}
    \includegraphics[width=0.4\textwidth]{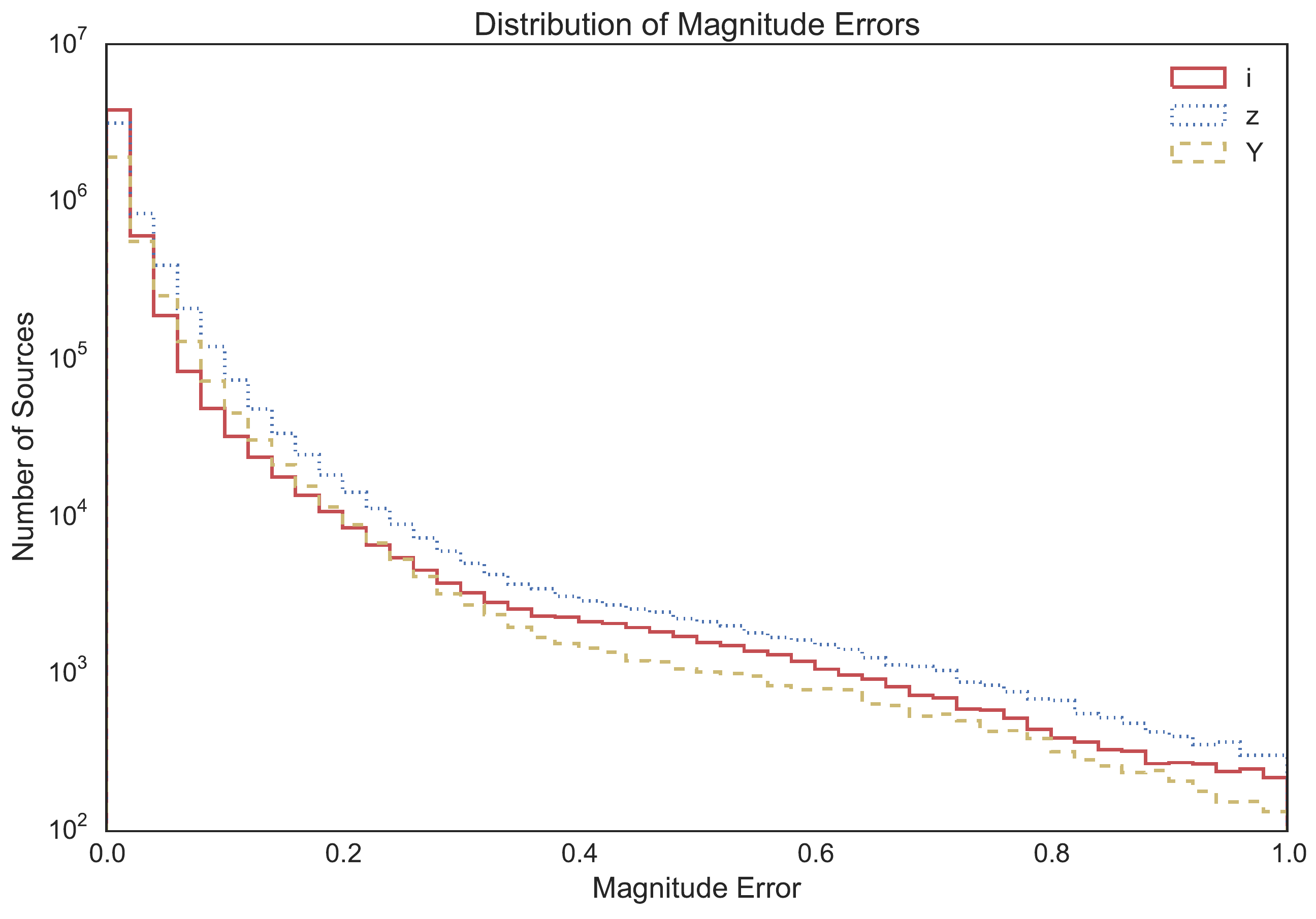}
    \includegraphics[width=0.4\textwidth]{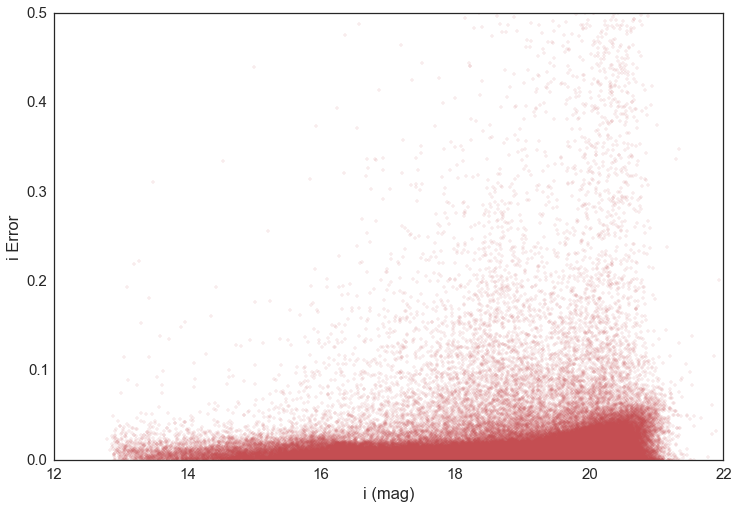}
    \caption{Distribution of magnitude errors for all 5+ million
      sources with observations in multiple bands (top) and $i$
      magnitude error as a function of magnitude for the F120 field
      (bottom). $\sim$78\% of $i$ band sources have better than 2\%
      photometry and $\sim$93\% have better than 5\% photometry (the
      mean error is $\sim$ 2\%). The errors in $zY$ are not quite as
      good, where both have $\sim$63\% of sources better than 2\%
      photometry and 79\% better than 5\% (with a mean of $\sim$3\%).}
    \label{fig:mag_error}
\end{figure}

To calculate the error for each source we look at the residual flux
left over by subtracting the PSF model (with the best fit parameters)
from the image patch inside the aperture. The error is then the ratio
of residual flux to PSF flux, which for the majority of our sources is
less than 2\%. Figure \ref{fig:mag_error} shows the percentage of
sources with PSF error $<$ 2\%, $<$5\%, and bad sources (with error
$>$5\%) for the entire SCOCENSUS catalog and for a field calibrated
with SDSS. By analyzing results on a select number of CCD images we
determined that nearly all the sources with errors $>$5\% are either
galaxies, sources with high background flux due to very bright distant
saturated stars, or unresolved binaries with slightly deformed shapes.

\subsection{Calibration} \label{scocensus-calibration}

\subsubsection{Photometric Calibration}

Once the images have been stacked, with instrumental PSF magnitudes
calculated, it is still necessary to calibrate the magnitudes to a
standard photometric system and calculate their proper motions. One
popular way to calibrate DECam photometry is to use a DECam native
system, fitting the stellar locus using a code like BIGMACS
\citep{Kelly2014}. This method was used by the DES in
\citep{Melchior2015} but requires $g$ band photometry to fit the
stellar locus, which we did not take.

\begin{figure*}\centering
    \includegraphics[width=0.9\textwidth]{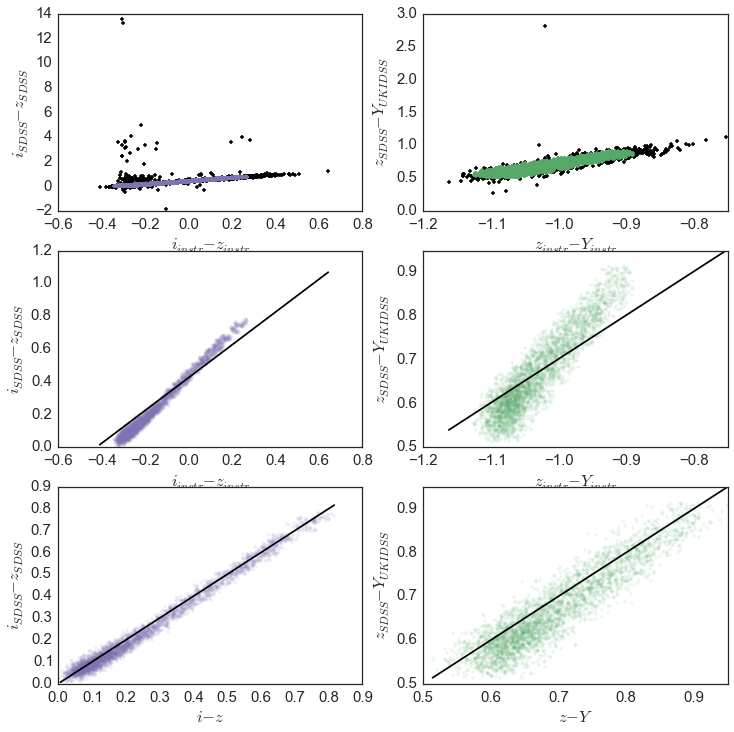}
    \caption{Corrections needed for $i-z$ colors (left column) and
      $z-Y$ colors (right column) to put DECam instrumental magnitudes
      on the SDSS AB photometric system (or $z_{SDSS}-Y_{UKIDSS}$ for
      $z-Y$). The top row shows all of the sources with good psf
      photometry with outliers removed before fitting the colors are
      shown in black. The middle row shows the difference between
      instrumental and SDSS magnitudes with a line with slope unity in
      black. The last row shows the same relationship after
      calibration, with a line with slope one in black.}
    \label{fig:color_correct}
\end{figure*}

Instead we calibrated our images by taking hourly exposures from
standard regions covered by SDSS (for $iz$ bands) and UKIDSS (for $Y$
band). Using the InstCal images we performed PSF photometry on all 3
filters and selected only the isolated, high signal to noise sources
(in 3 all bands) to use for calibration. Using \emph{astroquery}
\citep{Ginsburg2013} we loaded the relevant portions of the SDSS and
2MASS catalogs from
Vizier\footnote{\url{vizier.u-strasbg.fr/viz-bin/VizieR}}
\citep{Ochsenbein2000} and further restricted our calibration catalog
by choosing only sources SDSS flagged as single sources (i.e. no
binaries) with good stellar profiles. Unfortunately we neglected to
take standard fields in Stripe 82, which is well calibrated and has
flagged variable stars and other anomalous sources that could
interfere with our calibration. This causes our SDSS fields to have
several of sources with unusually large $i-z$ colors that we
reject. Once again we used MCMC to remove outliers but found another
method that was faster, in this case using the DBSCAN clustering
algorithm (\citealt{Ester1996}; see Figure \ref{fig:color_correct}).

\begin{figure*}\centering
    \includegraphics[width=0.9\textwidth]{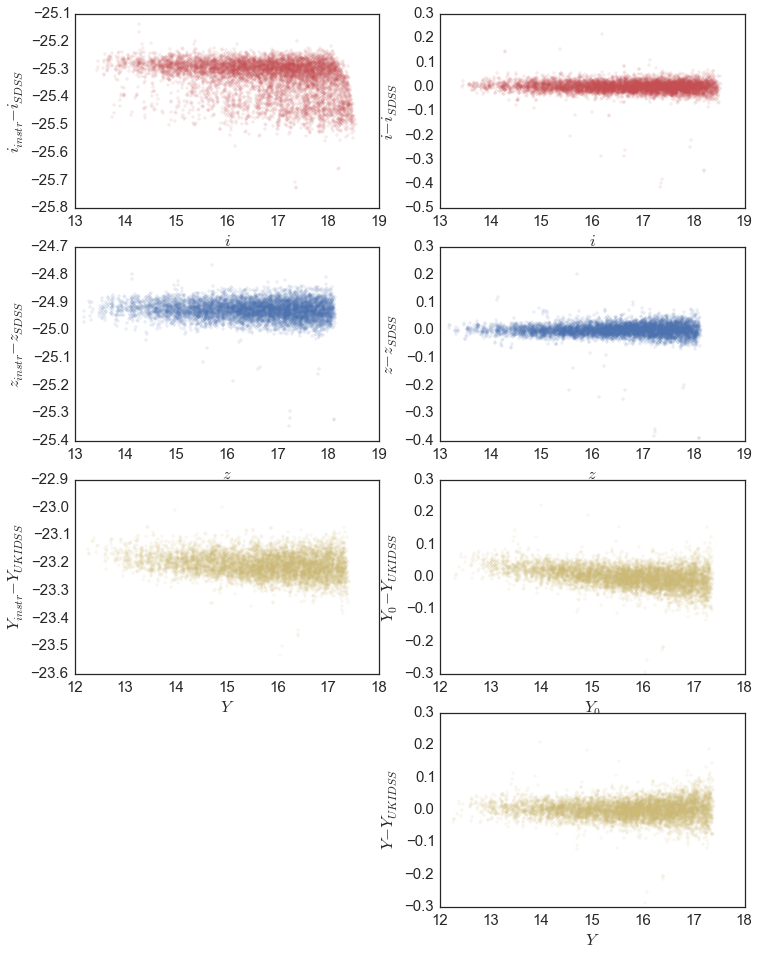}
    \caption{The left column shows the difference between DECam
      instrumental magnitudes and SDSS (and UKIDSS) magnitudes for
      matching sources. The right column shows the difference in
      magnitudes after calibration. In the case of the Y filter we
      show the difference without the extra magnitude-dependent term
      and with the magnitude-dependent term (bottom).}
    \label{fig:mag_correct}
\end{figure*}

While the DECam filters were designed to approximately match the SDSS
filters, slight differences in the total throughput and color
variations over the FOV make it necessary to add a color correction to
the photometric calibration (see Figure \ref{fig:color_correct}). To
calibrate our colors we use the transformations outlined in
\citet{Landolt2007}:

\begin{align}
    i_{SDSS}-z_{SDSS} &= a+b C_{iz} \\
    z_{SDSS}-Y_{UKIDSS} &= c+d C_{zY} 
\end{align}
where $a,b,c,d$ are coefficients used to make a linear transformation
between the reference colors and the instrumental color indices
$C_{iz},C_{zY}$ (colors outside the atmosphere), with

\begin{align}
    C_{iz} &= i_{instr}-z_{instr}-k_1 X - k_2 C_{iz} X \\
    C_{zY} &= z_{instr}-Y_{instr}-k_3 X - k_4 C_{zY} X
\end{align}
where $i_{instr}, z_{instr}, Y_{instr}$ are the instrumental
magnitudes, $k_x$ are the extinction coefficients, and $X$ is the
airmass.

This allows us to calculate the magnitude of each source in the SDSS system:

\begin{align}
    i &= i_{instr}-A_i X + Z_i + c_i (i_{SDSS}-z_{SDSS}) \\
    z &= z_{instr}-A_z X + Z_z + c_z (i_{SDSS}-z_{SDSS}) \\
    z_Y &= z_{instr}-A_{zY} X + Z_{zY} + c_{zY} (z_{SDSS}-Y_{UKIDSS})
\end{align}
where $i,z,Y$ are the calibrated magnitudes, $Z_x$ are the zero
points, $A_x$ are the extinction coefficients, $c_x$ are the color
coefficients, and $y_0$ is a magnitude dependent term $\sim$ 1.01 that
will be discussed shortly. Since some of the fainter (and redder)
sources may be detected in $z$ and $Y$ but not $i$ band images, we
calculate a second $z$ magnitude, $z_Y$, which has its colors
calibrated to the $z-Y$ color as opposed to the more accurate
$i-z$. To calculate the coefficients we use all but one of the SDSS
fields each night, using the last field as a test set to estimate our
photometric errors and verify that we are not overfitting the
data. Figure \ref{fig:mag_correct} shows the difference between our
magnitudes and the SDSS $iz$ magnitudes.

\begin{figure}\centering
    \includegraphics[width=0.45\textwidth]{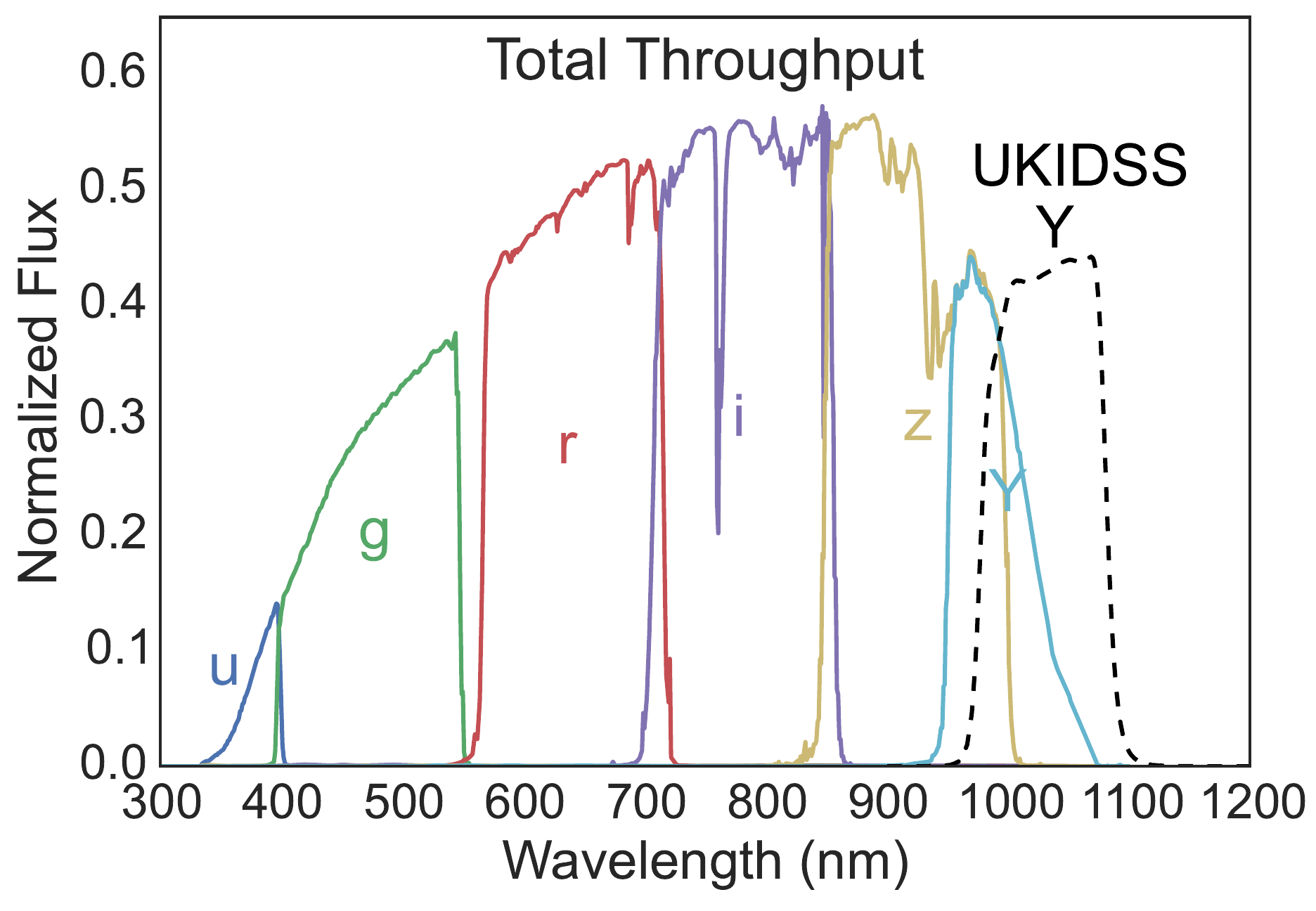}
    \caption{Total system throughput of the various DECam filters
      (solid lines) and the UKIDSS Y filter (dashed line). Note the
      extreme difference between DECam Y and UKIDSS Y.}
    \label{fig:throughput}
\end{figure}

\begin{figure}\centering
    \includegraphics[width=0.45\textwidth]{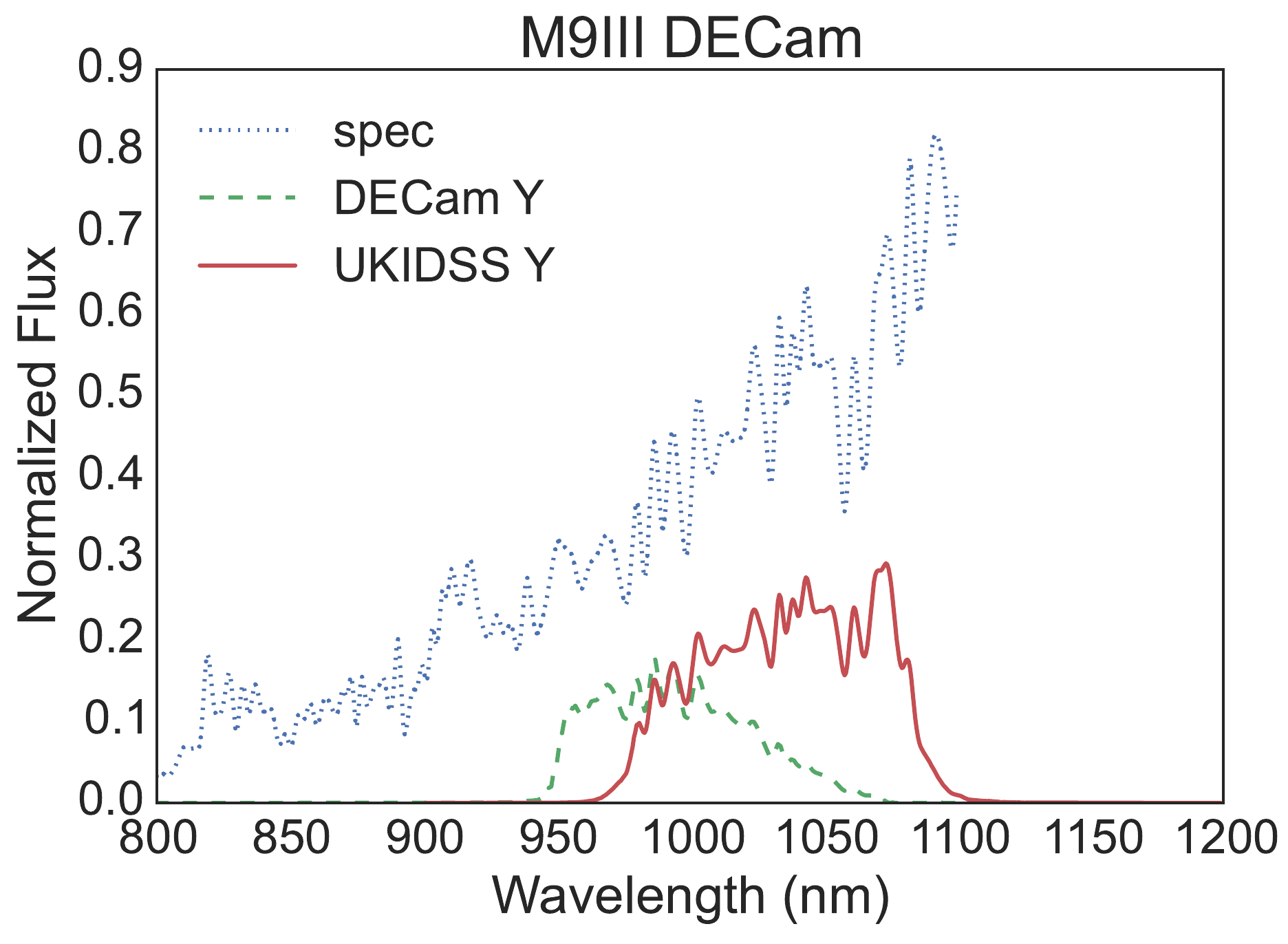}
    \caption{Sample M giant spectra from \citet{Pickles1998}. TiO
      bands and other features detected in UKIDSS Y and not DECam Y
      are likely the reason that our redder sources seem much fainter
      in the Y than UKIDSS even after calibration.}
    \label{fig:mgiant}
\end{figure}

\begin{figure}\centering
    \includegraphics[width=0.45\textwidth]{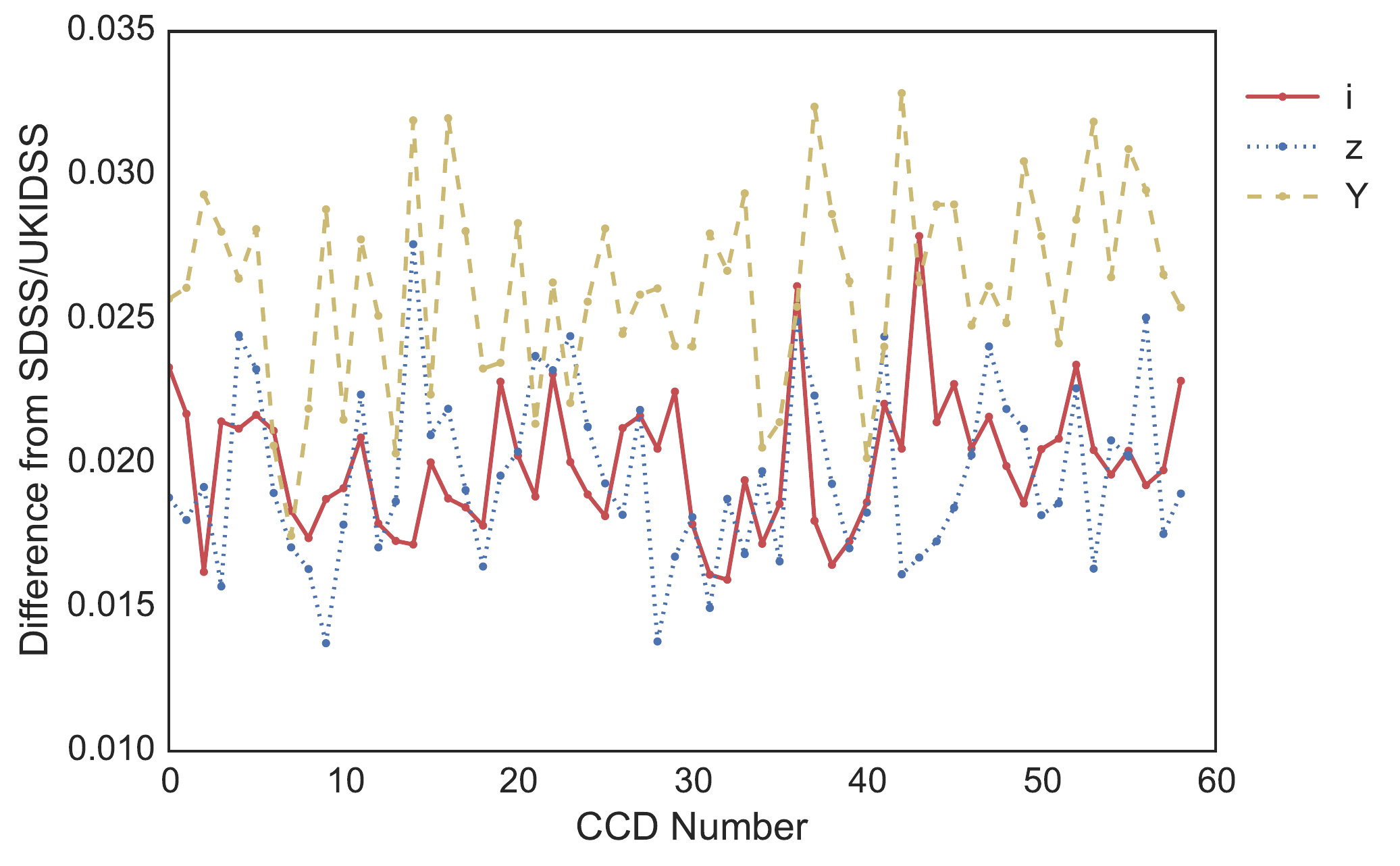}
    \caption{The RMS of the difference between our calibrated
      photometry and the corresponding magnitudes in SDSS (or UKIDSS
      for $Y$) for each CCD. Note: this ignores all sources where the
      difference in magnitude is $>0.1$ mag ($<$1\% of the total
      sources in the field), since our analysis has shown that those
      are sources that are either variable or have poorly constrained
      colors in SDSS.}
    \label{fig:sdss_rms}
\end{figure}

The remaining photometric task is to calibrate to UKIDSS $Y$, which is
not as straightforward as determining $i$ and $z$. Figure
\ref{fig:throughput} shows the throughput of the DECam filters and the
UKIDSS Y filter. Not only is the central wavelength of the two filters
different, the shapes of the filters are noticeably different and it
is likely that a color calibration is not likely to capture the
subtleties of differences in M dwarf spectra like TiO absorption (see
Figure \ref{fig:mgiant}). If we calibrate Y using similar coefficients
to $z_Y$ we notice that there appears to be a magnitude dependence in
our calibration (possibly due to a blue leak in the Y filter described
in \citealt{Hewett2006}). By modifying the $Y$ equation to
\begin{align}
    Y = M_0(Y_{instr}-A_Y X) + Z_Y + c_Y (z_{SDSS}-Y_{UKIDSS})
\end{align}
where $M_0 \sim 1.01$ is a magnitude dependent term, we see that our
calibration is much better but still less accurate than our $iz$
calibration. Properly modeling the difference between the DECam and
UKIDSS Y-band throughput might allow us to obtain finer calibrations
of our Y-band measurements but is not likely to produce enough of a
benefit (for our analysis) to justify spending additional time on
it. Figure \ref{fig:sdss_rms} shows the RMS for all of the CCDs for a
single night.

\subsubsection{Reddening and Extinction} \label{scoii-reddening}

\begin{figure*}\centering
    \includegraphics[width=.8\textwidth]{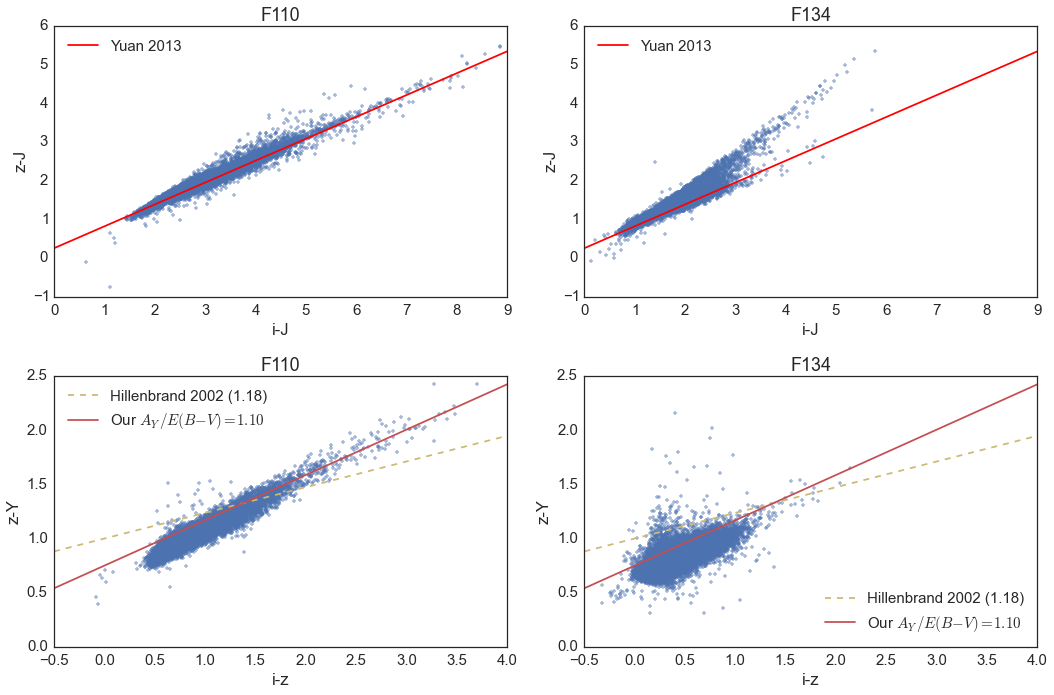}
    \caption{Comparison of reddening vectors calculated from
      $\frac{A_i}{E(B-V)}$, $\frac{A_z}{E(B-V)}$ and
      $\frac{A_J}{E(B-V)}$ from \citet{Yuan2013} and
      $\frac{A_Y}{E(B-V)}$ from \citet{Hillenbrand2002} with the
      reddened $\rho$ Ophiuchus field (F110) and a moderately reddened
      field in LCC (F134). While the \citet{Yuan2013} values match our
      observations, $\frac{A_Y}{E(B-V)}$ appears to be too high so we
      use the $\rho$ Ophiuchus field to estimate our
      $\frac{A_Y}{E(B-V)}=1.10$.}
    \label{fig:extinction}
\end{figure*}

To properly estimate effective temperatures (T${\textrm{eff}}$),
luminosity functions, and mass functions, we require knowledge of the
extinction in $izY$. \citet{Yuan2013} provides estimates for GALEX,
SDSS, 2MASS, and WISE passbands, including $A_i/E(B-V)=1.71$,
$A_z/E(B-V)=1.28$, and $A_J/E(B-V)=0.72$; while
\citet{Hillenbrand2002} estimates $A_Y/A_V=0.38$ for UKIDSS Y (1.035
$\mu$m). Using our most reddened field, $\rho$ Ophiuchus (F110), and a
moderately reddened field in LCC (F134), we measure the reddening
vectors $(A_z-A_J)/(A_i-A_J)$ using the extinctions predicted by
\citet{Yuan2013} and find good agreement with our observations (see
Figure \ref{fig:extinction}). Similarly we measure
$(A_z-A_Y)/(A_i-A_z)$, using $A_V/E(B-V)=3.1$ from \citet{Whittet1980}
to convert $A_Y/A_V$ to $A_Y/E(B-V)=1.18$, but find that it is
inconsistent with our observations. By fitting the red sources in a
$z-Y$ vs $i-z$ color-magnitude diagram (CMD), we estimate
$A_Y/E(B-V)=1.10$ (see Figure \ref{fig:extinction}).

\subsubsection{Astrometric Calibration and Proper Motions}
\label{scocensus-pmcal}

\begin{figure}\centering
    \includegraphics[width=0.4\textwidth]{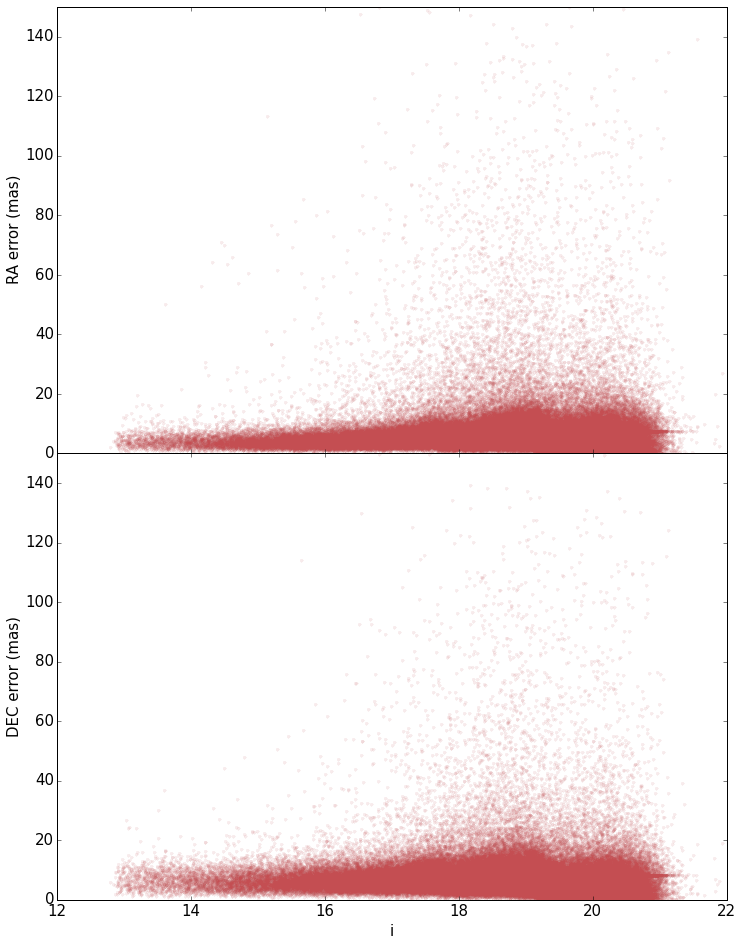}
    \caption{Positional errors in $izY$ in the final SCOCENSUS catalog
      in the F120 field. For all sources observed in 2015.4 the final
      position is the average windowed position extracted by
      \emph{SEP} from 6 different images: $izY$ stacks and $izY$ short
      exposures. The errors shown are the RMS values of the
      measurements used to determine the mean position. For sources
      only observed in a single image (for example faint sources only
      detected in $i$ stacks), the mean RMS is used. This is likely to
      underestimate the errors for sources observed in a single filter
      but does not affect our analysis as we require $iz$ colors for
      our candidate selection, meaning all of our sources have a
      minimum of two measurements for each position.}
    \label{fig:ra_err}
\end{figure}

To calculate mean positions and proper motions we use the same
procedure outlined in Section \ref{scocensus-astrometry} to reproject
exposures of the same field to a common reference frame. This time we
combine the short exposures and stacks in all 3 filters for a total of
6 possible exposures of the same source in a given epoch and calculate
the average of (x,y) image coordinates in the reference frame of the
z-band stacks (which we determined to have the lowest $\chi^2$ error
when projected onto the GAIA reference frame). This has to be done
separately for each CCD and each pointing. We still do not project the
$x$ and $y$ positions to a sky catalog yet, as we will reintroduce the
>100 mas errors that we have worked carefully to avoid. Instead, for
all of the fields observed in both 2013 and 2015, we reproject all of
the 2013 image coordinates to the 2015 image coordinates. This allows
us to estimate the relative proper motion of each source to within
$\sim$10 mas/yr, (20 mas position uncertainties with a two year
baseline). We then compare the proper motions we measured with sources
in GAIA with proper motion calculations and (after discarding
outliers) subtract the mean difference in RA and DEC, yielding an
absolute proper motion on the GAIA DR1 reference system.
Most likely due to the proximity of our fields to the Galactic plane,
less than 10\% of our observed sources have proper motions estimates
in the GAIA DR1 catalog \citep{GAIA2016}.

We also project our 2015 catalog to GAIA DR1 and match each source with
observations in the GAIA, UCAC4, 2MASS, USNOB 1.0, and
AllWISE catalogs. Although the position errors are
much larger than the precision of the DECam image coordinates, sources
detected in multiple catalogs with a large enough baseline can yield
reasonable proper motions (though still not as accurate as our proper
motions with both 2013 and 2015 observations). Figure \ref{fig:ra_err}
shows a sample of the internal RMS error for each source as a function
of magnitude for the F120 field after calibration. With an RA RMS of
11.0 mas, and DEC RMS of 11.3 mas, the precision of our catalog
positions is even better than the $\sim$20 mas precision of the DECam
CCDs, with 94.8\% of the sources having positional uncertainties less
than 20 mas in RA and DEC.

\section{Candidate Selection} \label{scoii-select-members}

Our candidate selection is made in several steps of color, magnitude,
and proper motion cuts to arrive at well-vetted lists of potential
members. The final cuts were chosen iteratively, where our initial
photometric cuts influenced our initial proper motion cuts, which
allowed us to choose better photometric cuts, etc., until we arrived at a
set of conditions that appear to eliminate the bulk of interlopers
while discarding as few members as possible.
This appendix describes in detail the final procedure used to select our candidates.

\subsection{Color-Magnitude Cuts} \label{scoii-cm-cuts}

\begin{figure}\centering
    \includegraphics[width=0.45\textwidth]{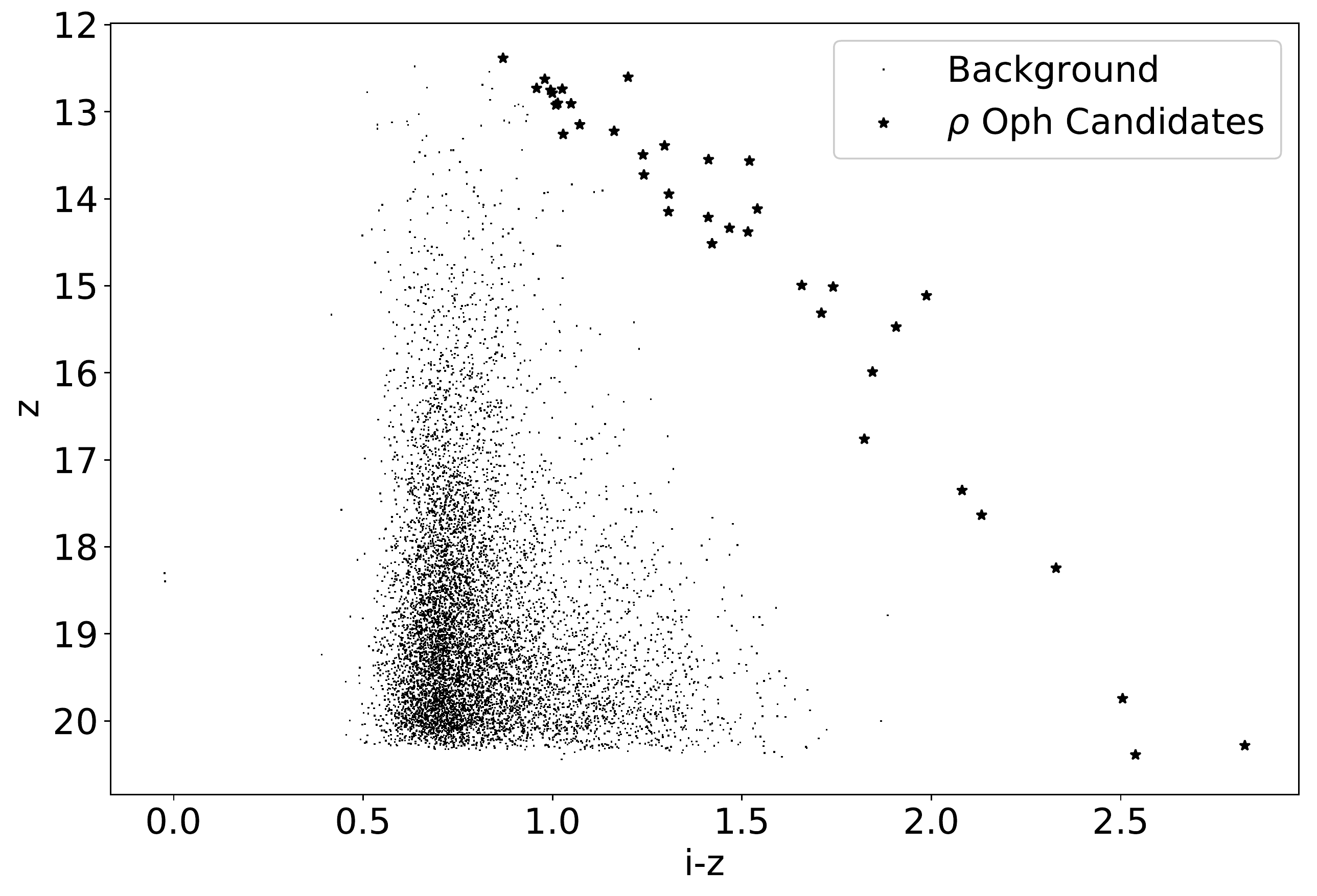}
    \caption{A subset of field F110 that covers the $\rho$ Ophiuchus star forming complex.
    We see that in a region with a small age and distance dispersion and nearly isotropic redenning,
    the members of the complex stand out from older background and foreground sources.
    The sources shown here are not included in this paper on UCL and LCC but will be
    analyzed in a future paper.}
    \label{fig:rho_oph}
\end{figure}

\begin{figure}\centering
    \includegraphics[width=0.45\textwidth]{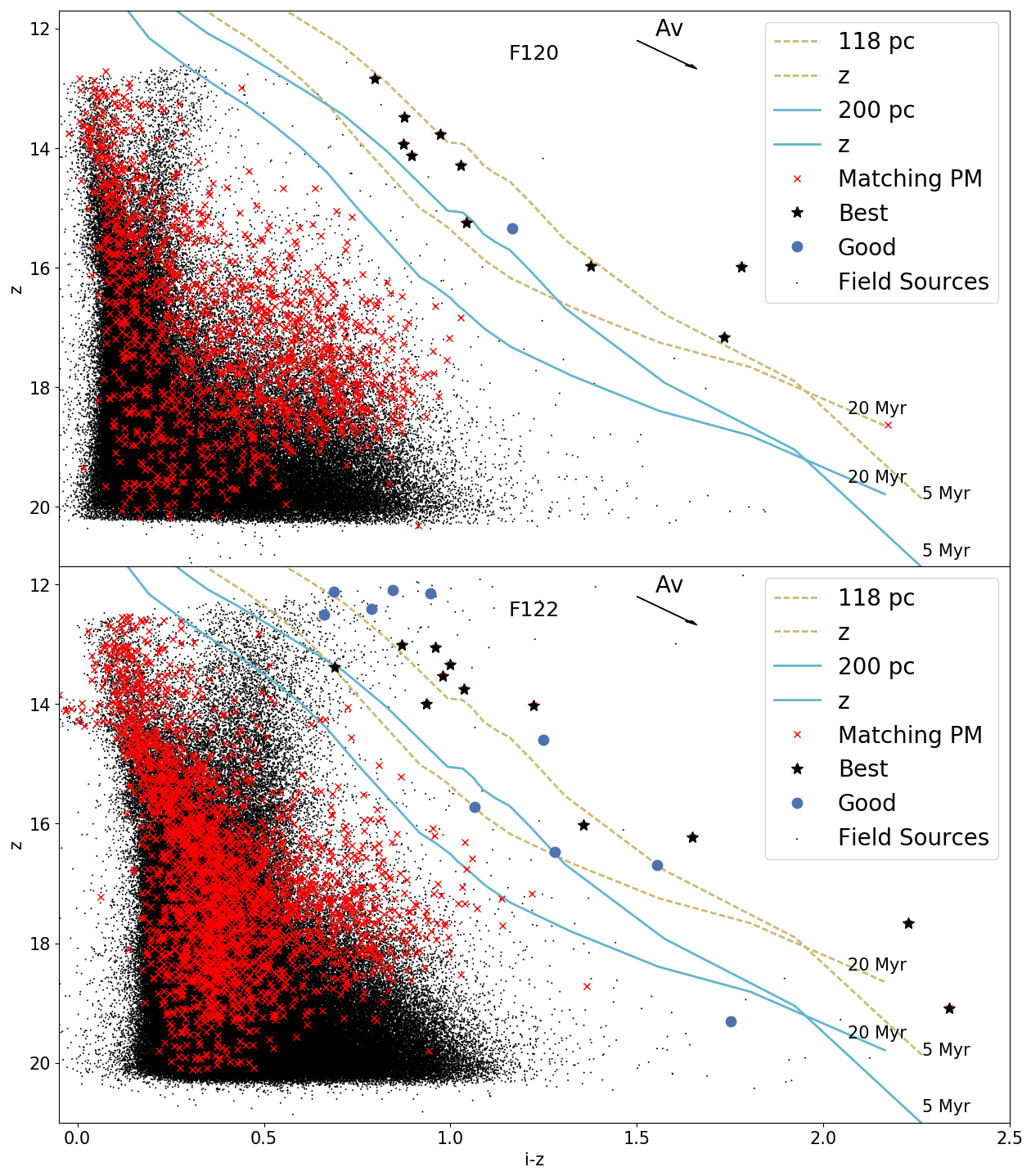}
    \caption{An example of an LCC field with negligible reddening
      (F120, top, with $b\sim12.5$) and one 6 degrees closer to the Galactic plane with
      significant reddening (F122, bottom, $b\sim6.6$). In F120 there is very
      little contamination from background sources between the 118 pc
      isochrones, where only a single object has proper motions
      consistent with LCC between the 200 pc isochrones. In the more
      reddened F122 there is a sizable number of interlopers for
      $z>14$ and 14 objects with similar proper motions between the
      200 pc isochrones. These are more likely to be interlopers than
      actual members and illustrate the necessity to choose a more
      stringent boundary for our candidates.}
    \label{fig:iso_limits}
\end{figure}

The first (and most stringent) set of cuts we make is selection based on $i-z$ vs $z$ CMD positions.
Figure \ref{fig:rho_oph} shows a subset of our survey that covers the $\rho$ Ophiuchus star forming region
(to be analyzed in a future paper), where we clearly see substellar sources tracing out an isochrone
in $i-z$ vs $z$ color-magnitude space.
Unfortunately the UCL and LCC fields investigated in this survey contain a lower density of YLMOs
and varying degrees of redenning in a single field of view
(even though most of our LCC and UCL objects show $A_\textrm{v}<0.1$, background sources can be
more strongly reddened).
This causes color-magnitude space to become polluted with reddened objects,
making it difficult to determine the isochrone for each field and select objects based on photometry alone
(see Figure \ref{fig:iso_limits}).

To make our color cuts we compare our models to the \citet{Allard2011} BT-Settl atmospheric models,
using the \citet{Asplund2009} abundances to estimate the effective temperatures (T$_\textrm{eff}$) of
our sources.
We create a linear interpolation function over the BT-Settl grid for
SDSS $iz$, UKIDSS $Y$, and 2MASS $JHK_s$ as a function
of T$_\textrm{eff}$ and log($g$) (surface gravity).
Because the observed magnitudes in the BT-Settl models are given at the stellar surface,
they depend on the radius of each stellar/substellar source,
which requires a set of isochrones for proper calibration.
For this we use the BHAC2015 \citet{Baraffe2015} stellar evolutionary tracks,
which allow us to match T$_\textrm{eff}$ and log($g$) from the BHAC2015 isochrones to
the BT-Settl atmospheric models.
We then use the absolute $M_J$ magnitudes from BHAC2015 to calibrate the magnitudes in
our BT-Settl grid, which can be converted into observed magnitudes by estimating the distance to
our sources.
It is regrettable that we require models for our photometric selection because,
as was mentioned in the introduction, it is known that these models
underestimate the flux for YLMOs at a given age.
This means that while the age of UCL and LCC sources are expected to range from $\sim11-25$ Myr
\citep{Pecaut2016}, our data is better approximated by isochrones from 5-20 Myr.
Since the distance to most UCL and LCC sources ranges from $\sim$100-200 pc \citep{Chen2011},
we use the 200 pc at 20 Myr isochrone as the lower magnitude limit for all sources with $i-z<1.7$.
We do not have an upper magnitude limit,
as sources redder than models predict are still likely to be interesting,
and we accept all objects with $i-z \geq 1.7$,
since the BHAC2015 models indicate that lower mass sources might become bluer and there are very few background
contaminants this red (see Figure \ref{fig:iso_limits}).
After applying these color cuts only $4.7x10^{-2}\%$ of the sources from our full catalog remain as
potential members.

\subsection{Kinematic Cuts} \label{scoii-pm-cuts}

Kinematic selection is performed by comparing our \emph{DECam} and \emph{Sky} proper motions
(as defined in Section \ref{scoii-obs}) with the average subgroup velocities predicted
by \citet{Chen2011} for UCL and LCC, for all of our sources with errors $\leq 20$ mas/yr,
giving us three different levels of cuts.
Sources observed in 2013 and 2015 that pass all photometric cuts and have both
\emph{DECam} and \emph{Sky} proper motions consistent with Sco-Cen are labeled \emph{best} sources
(these necessarily come from the F111, F112, F113, F120, F121, and F122 fields).
Sources matching either \emph{DECam} or \emph{Sky} proper motions (but not both)
are labeled \emph{good} sources.
Sources observed in a single epoch with no previous astrometric measurements,
or proper motion errors $>20$ mas/yr, are flagged as having unconstrained proper motions
and labeled as ``no pm''.
The majority of these are faint sources below the detection level in 2MASS, DENIS,
AllWISE, and \emph{Gaia}, so our faintest sources (spectral type L0 and below) only have proper
motions in the fields observed in both epochs, but even those can frequently have (relatively) large errors.

\begin{figure}\centering
    \includegraphics[width=0.48\textwidth]{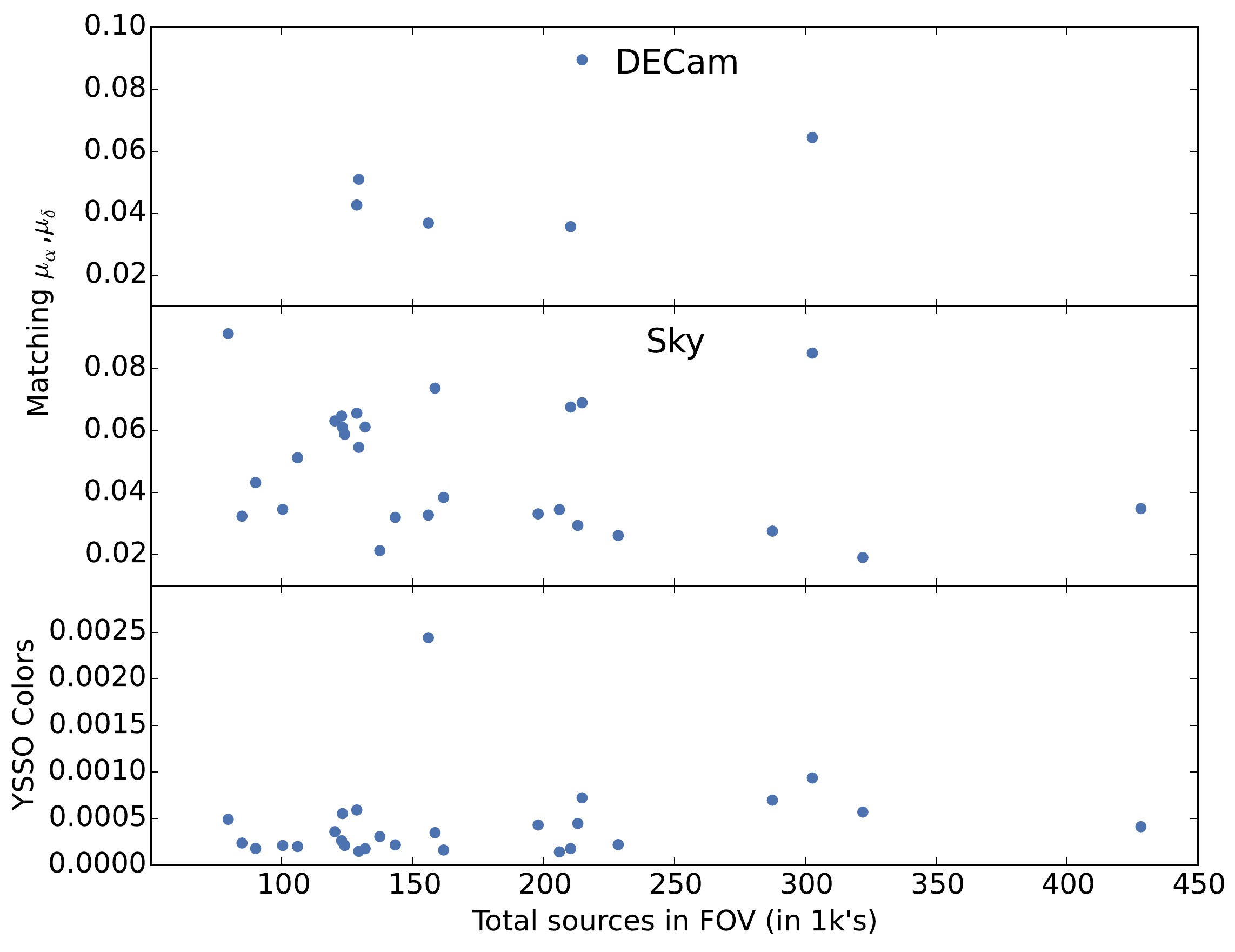}
    \caption{Fraction of all sources in each field with DECam proper
      motions (top), sky proper motions (middle), and photometry
      (bottom) consistent with YLMOs in UCL or LCC (depending on the
      field).}
    \label{fig:pm_frac}
\end{figure}

Figure \ref{fig:pm_frac} shows the fraction of all sources in each image that have proper
motions consistent with membership in Sco-Cen and the fraction of all
sources in each image that have $iz$ photometry consistent with membership in Sco-Cen.
Combining all of the fields together: 4.6\% of all objects have Sky proper motions consistent with Sco-Cen,
4.7$\times10^{-2}$\% have $iz$ photometry consistent with Sco-Cen,
and for objects observed in 2013 and 2015: 5.6\% have DECam proper motions
consistent with Sco-Cen and 0.96\% have both Sky and DECam proper
motions consistent with Sco-Cen.

\subsection{Photometric Model Fitting} \label{scoii-fitter}

Once we have applied the photometric and kinematic cuts we are left with 1680 candidate objects.
Using DECam $izY$ and 2MASS $JHK_s$ we can estimate T$_{\textrm{eff}}$,
giving us a spectral class template, by using the BT-Settl and BHAC2015
model grid described in Section \ref{scoii-cm-cuts}.
The easiest way to estimate T$_{\textrm{eff}}$ is to estimate distance, log($g$), age, and A$_\textrm{V}$,
and use a least squares-like algorithm to calculate the best fit T$_{\textrm{eff}}$.
This is problematic, as UCL and LCC are known to vary in age from $\sim11-25$ Myr \citep{Pecaut2016} and in
distance from $\sim100-200$ pc \citep{Chen2011}, so using a mean age of 16 Myr and distance of
118 pc for LCC and 142 pc in UCL results in degeneracies that lead to large errors in our T$_{\textrm{eff}}$
estimates.
Using spectral followups on a subset of candidates to be published in a future paper \citep{Moolekamp2018}
we compare the spectroscopic spectral types to our predictions obtained by fitting T$_{\textrm{eff}}$
and find that we can do a much better job estimating spectral types using a Monte Carlo Markov Chain (MCMC).

We use the emcee \citep{Foreman-Mackey2013} Markov Chain Monte Carlo (MCMC)
package to sample the same theoretical grid discussed above and produce estimates of
T$_{\textrm{eff}}$, $\log (g)$, distance, extinction (A$_\textrm{V}$), and mass.
In addition to fitting extra parameters, comparing the spectral templates estimated by our MCMC
algorithm shows much better agreement with our spectroscopic spectral types, so it is these
MCMC derived parameters that we include in our final catalog.

\subsection{Removing M Giants} \label{scoii-bye-giants}

Other than sources with unconstrained proper motions, the largest population of interlopers remaining in
our candidate list are M giants which have similar $izY$ colors but very different $JHK_s$ colors.
Removing these giant stars could have been performed before we fit the candidates to model photometry
by assuming A$_\textrm{V}\sim0.1$ for UCL and LCC, which \citet{Pecaut2016} show is a reasonable estimate.
But because A$_\textrm{V}$ is an output of our MCMC algorithm and the values we obtain appear reasonable,
we use the estimated A$_\textrm{V}$ for each source and the redenning vectors
described in Section \ref{scoii-reddening} to remove all sources more than 0.25 mag from
the ($H-K_s$,$J-H$) color-color positions predicted by the models.
Our giant star cuts are based on the 5 Myr isochrone because there is very little difference in color-color space
between the 5 and 10 Myr models and the 5 Myr isochrone contains photometry for lower mass objects, making it
easier to eliminate redder giant stars.
This is reasonable, as Figure \ref{fig:iso_limits} shows that many
of our sources lie on or above the 5 Myr isochrone.
Once all of our cuts have been applied we are left with a candidate list of mostly M dwarfs,
with a large number of fainter objects with unconstrained proper motions
that cannot be distinguished based on $izY$ colors alone,
leaving us with 562 (163 with proper motions) candidates in LCC
and 339 (234 with proper motions) in UCL.

\begin{figure}
    \includegraphics[width=0.48\textwidth]{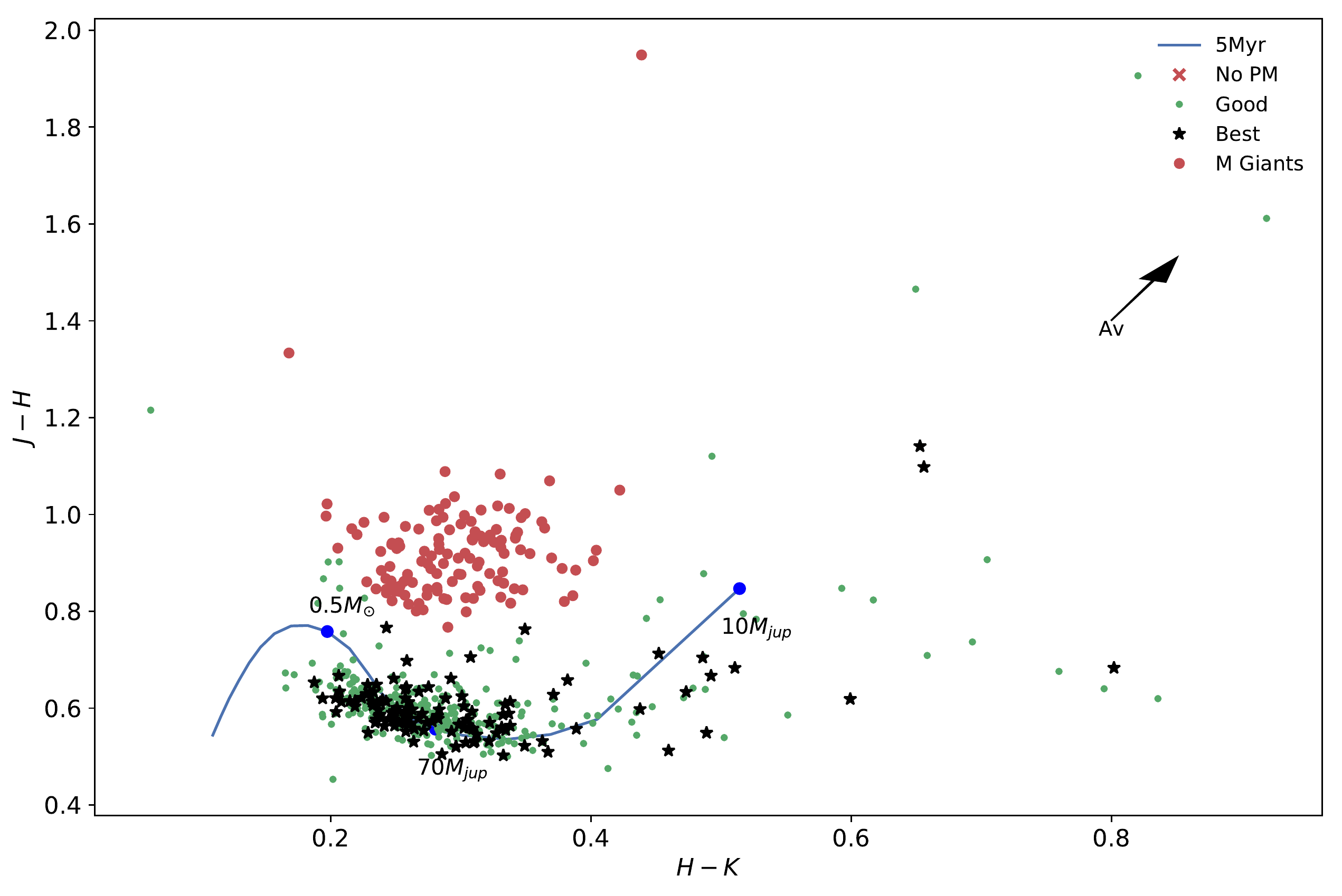}
    \caption{Color-color plot using 2MASS colors to separate M giants
      from M dwarfs. Sources with $J-H>0.25$ above the 5Myr isochrone
      are flagged as giants and eliminated. There are also a few
      sources with very strange $JHK_s$ colors ($H-K_s$<0 and $H-K_s>2$)
      that are not shown. The source flagged with no proper
      motion even though it has a 2MASS detection has a very high
      uncertainty in its proper motions ($>20$ mas) and is
      consistent with both Sco-Cen and zero proper motion.}
    \label{fig:mgiants}
\end{figure}

\begin{figure*}\centering
    \includegraphics[width=.9\textwidth]{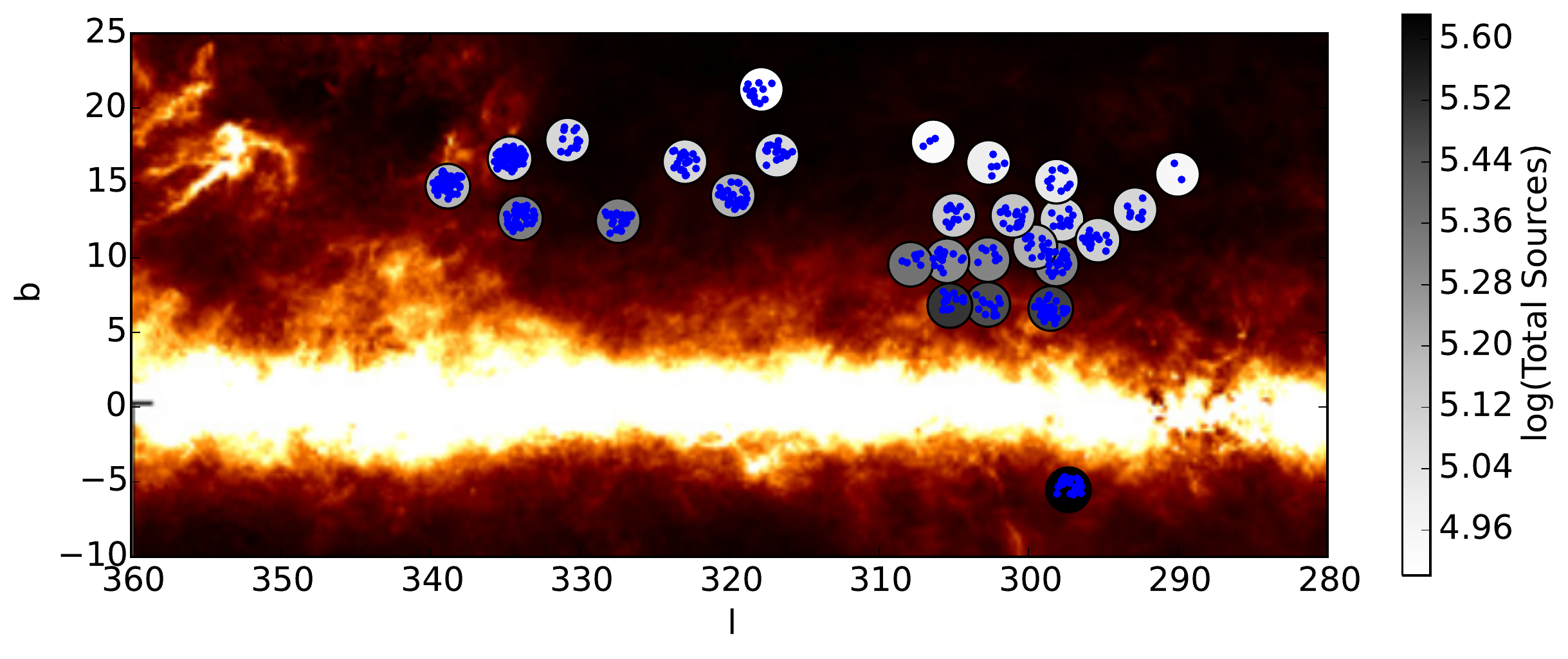}
    \caption{Location of new YLMO's presented in this paper, in
      Galactic coordinates, plotted over the SFD dust map
      \citep{Schlegel1998}. The blue dots are the locations of YLMOs
      and the large circles are centered at the location of each field
      of view (not to scale), colored to show the total number of
      objects in the field. While the total number of sources
      unsurprisingly increases closer to the Galactic plane, the
      number of YLMO candidates is more strongly correlated to the
      number of stellar sources in the field (see Figure
      \ref{fig:ylmo_stellar_ratio}).}
    \label{fig:ylmo_density}
\end{figure*}

\subsection{Estimating Contaminants} \label{scoii-selection-results}

\begin{figure}\centering
    \includegraphics[width=0.48\textwidth]{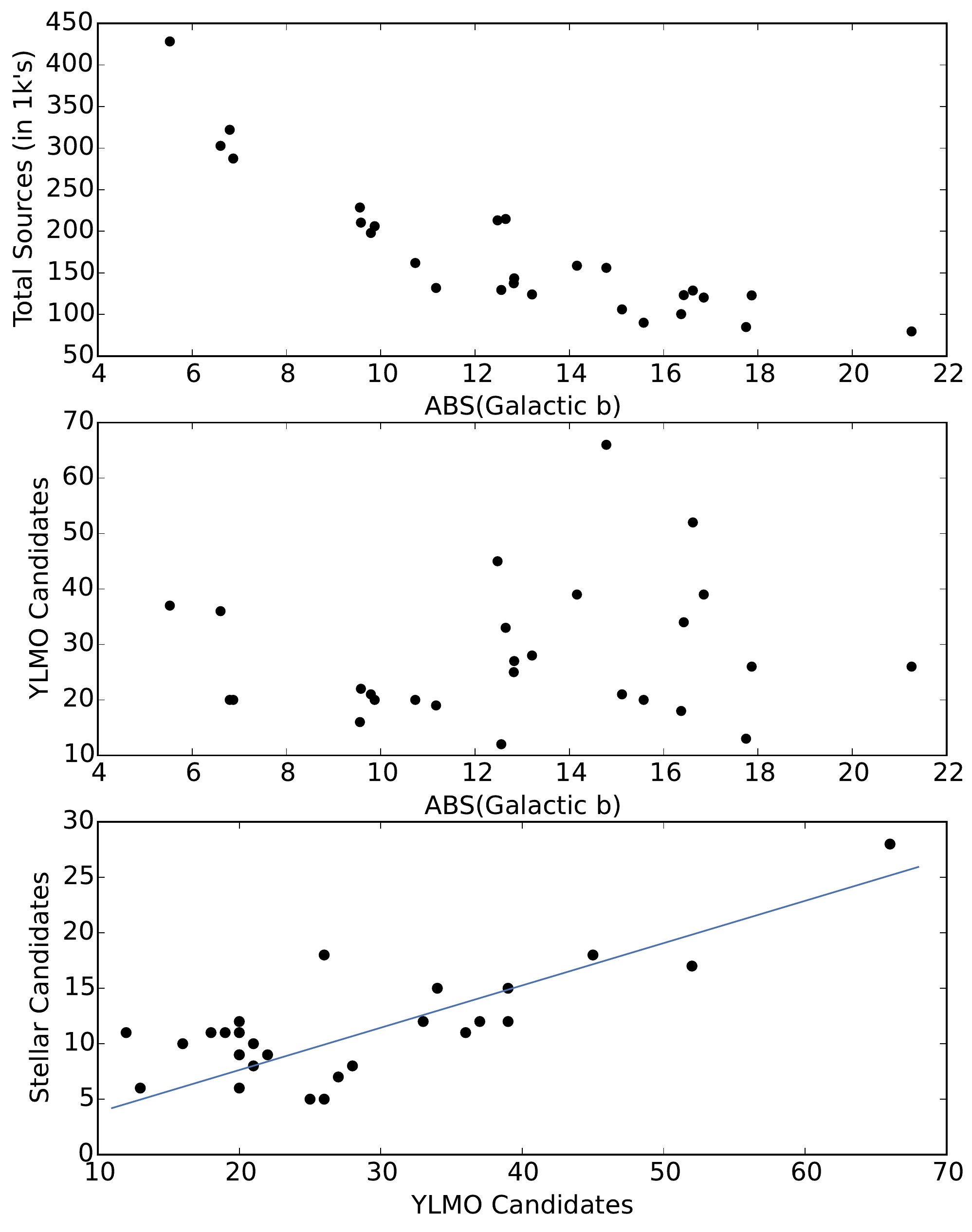}
    \caption{Comparison of the number of sources in a field and it's
      proximity to the Galactic plane (top), the number of YLMO
      candidates and their proximity to the Galactic plane (middle),
      and the ratio of higher mass stellar candidates/members and YLMO
      candidates in each field. While the total number of objects in a
      field increases as it approaches the Galactic plane (top plot),
      the total number of YLMOs in a field does not (middle
      plot). Instead there is a much stronger correlation between the
      number of stellar candidates in a field and the number of YLMO
      candidates, where a line with $\frac{\textrm{number of
          YLMOs}}{\textrm{number of high mass stars}}=2.62$ is shown
      (bottom plot).}
    \label{fig:ylmo_stellar_ratio}
\end{figure}

We saw in Figure \ref{fig:iso_limits} that fields near the Galactic
plane can have a substantial number of reddened sources that appear
photometrically similar to objects in Sco-Cen and it is necessary to
ensure that they aren't polluting our candidate list. Figure
\ref{fig:ylmo_density} shows that the number of sources in a field
increases as it approaches the Galactic center, while the number of
candidates in a field does not. This is displayed more quantitatively
in Figure \ref{fig:ylmo_stellar_ratio}, where the number of YLMO
candidates in a field is shown to be correlated with the number of
higher mass stellar candidates in a field (roughly 8:3) and not the
fields proximity to the Galactic plane.
This gives us a preliminary check that our pipeline is correctly selecting
candidates with minimal background contamination.

We also used spectroscopic follow-up of 17 objects performed using the ARCoIRIS spectrograph last
summer to verify both the spectral class of our objects and signatures of low surface gravity that
indicate young objects (as opposed to older main-sequence stars or M giants).
Of the 17 observed objects, only 1 of them appears to be an older star.
These objects were taken from both our ``best'' and ``good'' candidate lists, so we estimate
that $\sim5-6$\% of our objects are likely to be interlopers.
This estimate is also supported by our calculation that 5.6\% of our sources have proper motions consistent
with Sco-Cen, most of which are coincidental since $<<1\%$ are photometrically consistent
(see Section \ref{scoii-pm-cuts}).
This is only a rough estimate, as our brighter sources are likely to contain contamination due to
reddened stars in the Galactic plane and our fainter substellar sources have larger errors in their
proper motions.
In the future, creating a model for reddened stars in the Galactic disk will give us a better
understanding of the photometric contamination and allow us to provide
a more accurate estimate of our background contamination.


\bsp	
\label{lastpage}
\end{document}